\documentclass{aa}  
\pdfoutput=1 
\usepackage{hyperref}
\usepackage{graphicx}
\usepackage{txfonts}
\usepackage{amsmath}
\usepackage{float}
\usepackage{siunitx}
\usepackage[dvipsnames]{xcolor}
\usepackage{physics}
\usepackage{pdflscape}
\usepackage{float}
\usepackage{afterpage}
\usepackage{csquotes}
\usepackage{nameref}
\usepackage{pdflscape}

\begin{document} 



\title{First measurement of the triaxiality of the inner dark matter halo of the Milky Way}


\author{{Hanneke C. Woudenberg} 
          \and 
          {Amina Helmi}
          }

   \institute{Kapteyn Astronomical Institute, University of Groningen, Landleven 12, NL-9747 AD Groningen, the Netherlands,}

   \date{Received ..... .., ....; accepted ......, ....}

\abstract
  {Stellar streams are particularly sensitive probes of the mass distribution of galaxies.}
   {In this work, we focus on the Helmi Streams, the remnants of an accreted dwarf galaxy orbiting the inner Milky Way. We examine in depth their peculiar dynamical properties, and use these to provide tight constraints on the Galactic potential, and specifically on its dark matter halo in the inner 20 kpc.}
   {We extract 6D phase-space information for the Helmi Streams from {\it Gaia} DR3, and confirm that the Streams split up into two clumps in angular momentum space, and that these depict different degrees of phase-mixing. To explain these characteristics we explore a range of Galactic potential models with a triaxial NFW halo, further constrained by rotation curve data.}
  {We find that a Galactic potential with a mildly triaxial dark matter halo, having $p=1.013^{+0.006}_{-0.006}$, $q=1.204^{+0.032}_{-0.036}$, $M_{\rm{discs}}=4.65^{+0.47}_{-0.57}\cdot10^{10} M_{\odot}$ and $M_{\rm{DM}}(< 15 \text{kpc})=1.14^{+0.11}_{-0.10}\cdot10^{11} M_{\odot}$, is required to form two clumps in angular momentum space over time. Their formation is driven by the fact that the clumps are on different orbital families and close to an orbital resonance.  This resonance also explains the different degrees of mixing observed, as well as the presence of a dynamically cold subclump (also known as S2).}
   {This first and very precise measurement of the triaxiality of the inner dark matter halo of the Galaxy uniquely reveals the high sensitivity of phase-mixed streams to the exact form of the gravitational potential.}


\authorrunning{H.C. Woudenberg and A. Helmi}
\titlerunning{A first measurement of the Milky Way's inner dark matter halo's triaxial shape}

\maketitle
\section{Introduction} \label{sec:intro}

Structure formation is hierarchical in the $\Lambda$CDM cosmological model, with cold dark matter (CDM) halos predominantly growing via mergers \citep[e.g.][]{WhiteRees1978}. For Milky Way-like galaxies, this usually involves a small number of major mergers at early times and many minor mergers throughout their history \citep[e.g.][]{Cooper2010, Fatthai2020}. While in CDM only simulations DM halos have triaxial shapes \citep{Frenk1988, Springel2004}, in hydrodynamical simulations including baryonic physics the DM halos tend to be rounder due to presence of baryons  \citep{Chua2019, Prada2019}. In general DM halos are more spherical in the inner regions and are found to have radially varying shapes \citep{Zavala2019, Shao2021}. Further, the evolution of a DM halo's shape is influenced by its merger history and its mode of accretion \citep{VeraCiro2011DMhalo, Cataldi2021}, and also by the fundamental nature of the DM particle \citep{Vargya2022}.

The merger history of the Milky Way has been studied in detail over the past decades, starting with the discovery of the Sagittarius dwarf galaxy \citep{Ibata1994} and its tidal tails \citep{Yanny2000, Ivezic2000}, and the discovery of the nearby Helmi Streams \citep{Helmi1999}. By now, we have established a picture of the Milky Way's merger history which seems to have been rather quiescent. The advent of \textit{Gaia} DR2 revealed that the last major merger of the MW happened about 10~Gyr ago with the massive dwarf galaxy \textit{Gaia}-Enceladus-Sausage \citep[GES,][]{Helmi2018, Belokurov2018}. The debris of a range of smaller dwarf galaxies has also been found in the inner Galactic halo \citep[e.g.][]{Koppelman2019halo, Myeong2019, Naidu2020, Forbes2020, Horta2021, Dodd2023}. These accreted dwarf galaxies also bring in their own globular clusters and possibly satellite galaxies \citep{Massari2019, Kruijsen2019, Horta2020, Callingham2022, Malhan2022}. Furthermore, more than 100 thin stellar streams have been discovered so far \citep{Carlberg2018, Bonaca2021, Mateu2022, Malhan2022}. 

Such narrow distant stellar streams have been used to study the mass distribution of our Galaxy and particularly the shape of its DM halo. Examples include the modelling of GD-1 \citep{Koposov2010, Bovy2016GD1Pal5}, the disrupting Palomar 5 \citep{Kupper2015} and the Sagittarius Stream \citep{LawMajewski2010, VeraCiro2013, VasilievSgr2020}. In recent years it has become clear that the determination of the shape of our DM halo is further complicated by the perturbations from, for example, the Large Magellanic Cloud \citep[LMC, e.g.][]{VasilievTango2021, Vasiliev2023_LMC}. 

Some authors have recently pointed out that non-integrability of the Galactic potential can also leave imprints on the morphology and dynamics of streams, especially on streams near or on separatrices, the transitions between orbit families \citep{PriceWhelan2016, Mestre2020, Yavetz2021, Yavetz2022}. Such streams provide us with a direct probe of the location of resonances, chaotic regions and the orbital families that are present in a gravitational system. These depend both on the characteristic parameters of the gravitational potential, such as the halo flattening, halo scale radius or disc mass \citep{Caranicolas2010, Zotos2014}, and also on the distribution function \citep{Valluri2012}.

In this work, we use the detailed dynamics of the Helmi Streams to constrain the Milky Way's gravitational potential, and in particular the shape of its DM halo. The Helmi Streams, the first ever substructure identified to occupy the inner Galactic halo \citep{Helmi1999}, are the remnants of an accreted dwarf galaxy. Follow-up studies over the past 25 years, starting with \cite{Chiba2000}, have created a picture of the history and characteristics of the Streams \citep[see also][]{Kepley2007, Smith2009}. The parent dwarf galaxy is thought to have a mass of about $10^8 \: M_{\odot}$ and to have been accreted 5 to 10~Gyr ago \citep{Kepley2007, Koppelman2019Helmi, Naidu2022, RuizLara2022HS}. The Helmi Streams' stars show abundance patterns that are distinct from other substructures \citep{Aguado2021, Nissen2021, Matsuno2022, Horta2023, Zhang2024}. Between 7 to 15 GC are thought to be associated to it \citep{Koppelman2019Helmi, Massari2019, Kruijsen2020, Forbes2020, Callingham2022}. As was noted by \cite{Helmi2020}, it seems curious that such a low mass galaxy would have been accreted to the inner halo at $z < 1$ and is now orbiting the inner Milky Way. Possible mechanisms could involve group infall, but as we shall demonstrate below, this might not be necessary and a purely dynamical explanation is likely. 

\cite{Dodd2022} found that the Helmi Streams split into two clumps in angular momentum space which are chemically indistinguishable, supporting a common origin. One of the clumps was found to exhibit a kinematically cold subclump \citep{Myeong2018, Dodd2022} Later, the two clumps were recovered using a single-linkage clustering algorithm in \textit{Gaia} EDR3 \citep{Lovdal2022, RuizLara2022} and \textit{Gaia} DR3 data \citep{Dodd2023}, with a gap identified in between as an underdensity. \cite{Dodd2022} used the existence of the gap to constrain the flattening of the DM halo density to be $q_{\rm{\rho}} = 1.2$. In such a potential, the gap is long-lasting as the effective gravitational potential (provided by the sum of mass all components) is close to being spherical, and one of the clumps is on an orbital resonance. This however does not explain how the two clumps were formed, which is the question this paper will address. We do this by inspecting the orbital structure in the region of phase space occupied by the streams for a range of Galactic potential models with a triaxial DM halo. 

This paper is structured as follows.  
Sect.~\ref{sec:datamethod} discusses the data and the methods used in this paper, which involves the application of a chaos indicator, the study of orbit families and orbital frequency analysis. 
In Sect. \ref{sec:analysis} we present our analysis of the orbital properties of the Helmi Streams in a range of triaxial potentials. This leads to a method to constrain the Galactic potential in the region probed by the streams, and in particular the shape of the inner DM halo. 
In Sect. \ref{sec:results}  we discuss the results of this analysis. 
We end with a general discussion in Sect.~\ref{sec:disc} and present our conclusions in Sect.~\ref{sec:conclusion}. 
\section{Data and Method} \label{sec:datamethod}

\subsection{Generalities}
\label{sec:datamethod:generalities}

We use the Helmi Streams (HS) sample of \cite{Dodd2023}, which was constructed by applying a single-linkage clustering algorithm \citep[see also][]{Lovdal2022, RuizLara2022} on the \textit{Gaia} DR3 \citep{GaiaDR3} RVS sample for a selection of halo stars in a local volume of 2.5~kpc. This sample consists of 319 members with distance errors~<~20\%. The distances to the stars were obtained by inverting their parallaxes after correcting these for a zeropoint offset following \cite{Lindegren2021}. 

To convert to Galactocentric cartesian and cylindrical coordinates\footnote{We use a right-handed Galactocentric cartesian coordinate system and denote the coordinates as $(x, y, z, v_x, v_y, v_z)$. The GC is at its centre, $x$ increases in the disc plane away from the location of the Sun, meaning the Sun is located at negative $x$, $y$ points in the direction of Galactic rotation and $z$ points towards the North Galactic Pole. Galactocentric cylindrical coordinates are denoted as $(R, \phi, z, v_R, v_{\phi}, v_z)$.}, we follow the assumptions of \cite{Eilers2019} and \cite{Zhou2023}, as we will use their rotation curve data later in this work, which requires consistency. Hence, we assume  a height above the midplane of $z_{\odot} = 0.025~\si{\: kpc}$ \citep{Juric2008} and a distance from the Sun to the Galactic Centre of $R_{\odot} = 8.122~\si{\: kpc}$ \citep{Gravity2018}, which is consistent with more recent measurements of orbits around Sgr~A$^*$ \citep{Do2019, GRAVITY2021} and other dynamical measurements \citep[e.g.][]{Leung2023}. To correct for the motion of we Sun, we use the proper motion of Sgr~A$^*$ by \citep{Reid2004}, $\mu_{\rm{l,Sgr \: A^*}} = -6.379 \pm 0.026$~mas~yr$^{-1}$, $\mu_{\rm{b,Sgr \: A^*}} = -0.202 \pm 0.019$~mas~yr$^{-1}$, and the relation that 
\begin{equation}
    v_j = 4.74057 \: \si{km \: s^{-1}} \left( \dfrac{\mu_j}{\si{mas \: yr^{-1}}} \right) \left( \dfrac{d}{\si{kpc}} \right)
    \label{eq:vel}
\end{equation}
where $j = (l,b)$ in this case. This gives $v_{y, \odot} = 245.6 \si{km \: s^{-1}}$ and $W_{\odot} = v_{z, \odot} = 7.8 \: \si{km \: s^{-1}}$. We assume $U_{\odot} = v_{x, \odot} = 11.1 \: \si{km \: s^{-1}}$ \citep{Schonrich2010}, which is consistent with a range of more recent estimates \citep{Tian2015, Mroz2019, Zbinden2019}. The $z$ component of the angular momentum vector, $L_z$, is defined such that it is positive for prograde orbits, i.e.~its sign is flipped. 

\begin{figure*}[htb!]
\centering
    \includegraphics[width=0.8\hsize]{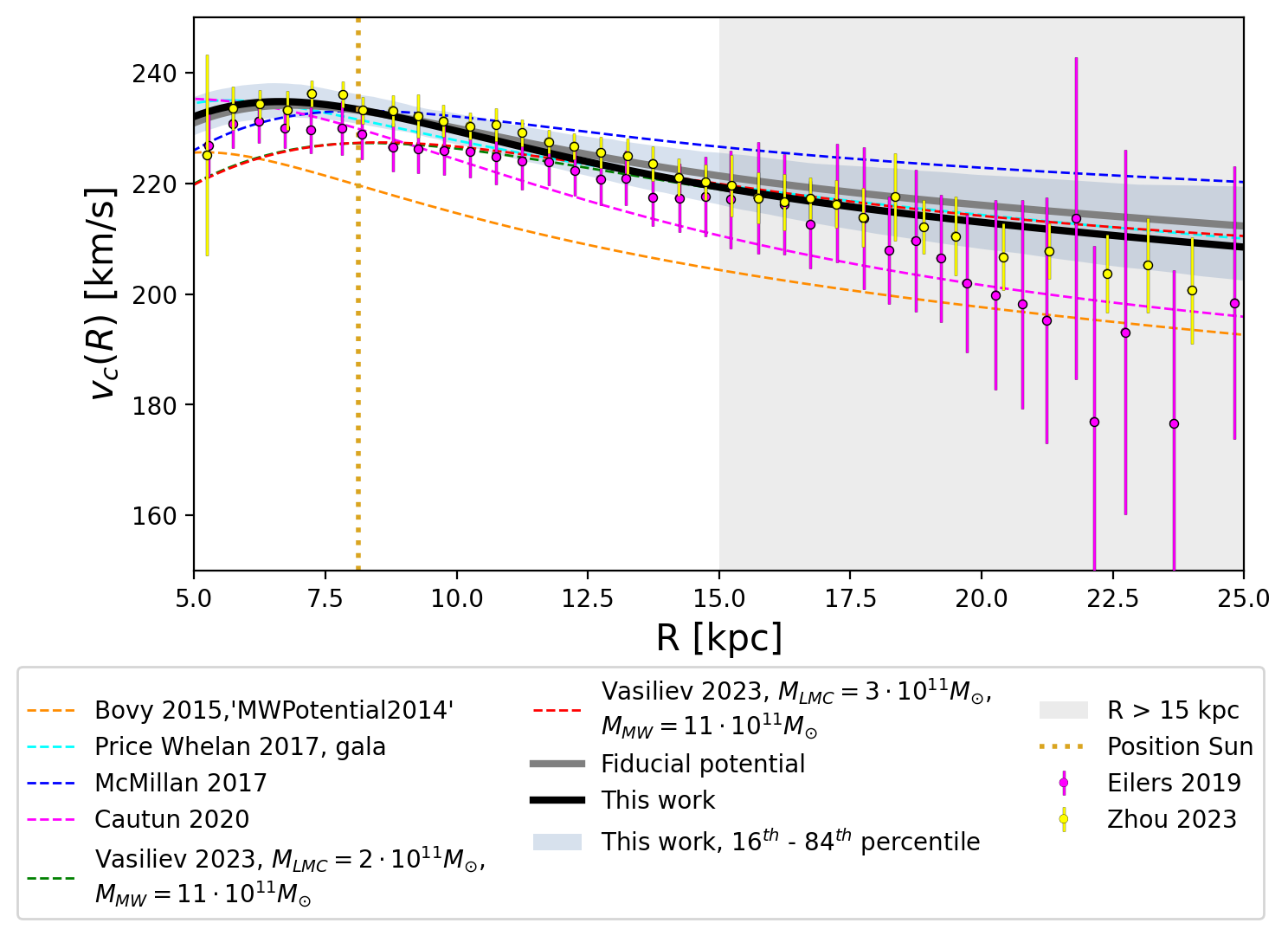}
\caption{Rotation curves of a variety of Milky Way mass models. We have plotted here the models by \cite{Bovy2015galpy}, \cite{softwarecitegala}, \cite{McMillan2017}, \cite{Cautun2020} and two models by \cite{Vasiliev2023_LMC} using his $\mathcal{L}3, \mathcal{M}11$ and $\mathcal{L}2, \mathcal{M}11$ halo models at the present day. The grey line shows the rotation curve of the fiducial model (described in Sect.~\ref{sec:datamethod:generalities} and Appendix~\ref{app:pot}). The black line shows the best-fit potential of this work (described in Sect.~\ref{sec:results:results}), while the grey shaded area correspond to the 16$^{th}$ to 84$^{th}$ percentile range for 200 potentials from the MCMC chains. The rotation curve data used in the fit includes \cite{Eilers2019}'s rotation curve data, shown in magenta, and \cite{Zhou2023}'s in yellow. The light grey shaded area marks the region $R>15$~kpc and data in this distance range was not included in the MCMC fit. However, the fit is similar even if data up to $R\sim 20$~kpc is used.
\label{fig:RC_fit}}
\end{figure*}

We use \texttt{AGAMA} \citep{AGAMA} for orbit integration and to compute dynamical quantities. Our fiducial MW model follows the \cite{McMillan2017} potential with some modifications. The \cite{McMillan2017} axisymmetric potential model consists of a bulge, an exponential thin and thick disc, an HI gas disc, a molecular gas disc and a spherical NFW DM halo. Instead of the 3~kpc of the original model, in this work, we  set a thick disc scale length of $R_{\rm{thick}} = 2$~kpc\footnote{A value of $R_{\rm{thick}}= 2$~kpc is in agreement with the findings of \cite{Bensby2011} and \cite{Bland-Hawthorn2016} and references therein. Such a thick disc thus has a smaller scale length than the thin disc, $R_{\rm{thin}} = 2.5 $~kpc, as we could expect from their formation histories \citep[see e.g.][]{Villalobos2010, Xiang2022}.}. We also assume a prolate DM halo with a flattening in the density of $q_{\rm{\rho}} = 1.2$, following \cite{Dodd2022}. To constrain the mass of the stellar discs, the DM halo scale radius and density in this new fiducial model, we use recent rotation curve data with associated uncertainties derived using axisymmetric Jeans equations by \cite{Eilers2019} and \cite{Zhou2023}. The procedure followed to determine the values of the characteristic parameters is described in detail in Appendix~\ref{app:pot}. The resulting rotation curve and a comparison to other models from literature is shown in Fig.~\ref{fig:RC_fit}. Our fiducial model has a total mass within 20~kpc of $M_{\rm{tot}}(r < 20 \: \si{kpc}) = 2.2^{+0.1}_{-0.1} \cdot 10^{11} M_{\odot}$, compatible with other estimates \citep[e.g.][]{Kupper2015, Posti2019, Watkins2019, Eadie2019}, and $v_{\rm{c}}(R_{\odot}) = 233.2 \pm 2.6 \: \si{km \: s^{-1}}$. Consequently, our $V_{\odot} = v_{y, \odot} - v_{\rm{LSR}} = 12.4 \: \si{km \: s^{-1}}$, well in agreement with for example the work by \cite{Tian2015}, \cite{Zbinden2019} or the commonly used value found by \cite{Schonrich2010}. 

\subsection{Review of dynamics}
\label{sec:review}

Integrals of motion (IoM) are quantities that only depend on a body's phase-space coordinates and are constant along an orbit. In a time independent potential, the total energy $E = E_{kin} + E_{pot}$ is an IoM. In a spherical potential, all components of the angular momentum vector, $\vec{L} = \vec{r} \cross \vec{p} = (L_x, L_y, L_z)$, are IoM, and hence so is the perpendicular angular momentum vector component, $L_{\bot}^2 = L_x^2 + L_y^2$. In an axisymmetric potential, symmetry with the polar angle $\theta$ is broken and therefore only $L_z = R v_{\phi}$ remains an IoM, though $L_{\bot}$ is sometimes used to characterise orbits (as a proxy for a third integral). In a triaxial potential, also symmetry with respect to the azimuthal angle $\phi$ is broken and none of the components of the angular momentum vector are IoM. 

A regular orbit is quasi-periodic, and the time series of its phase space coordinates $w(t) = \left(\mathbf{x}(t), \mathbf{v}(t)\right)$ can approximated by the sum
\begin{equation}
    w(t) = \sum_{k}^{\infty} a_k e^{i \omega_k t},
    \label{fourier}
\end{equation}
\noindent where $\omega_k$ are orbital frequencies and $a_k$ the associated amplitudes. Therefore, a Fourier transform of the orbit will give us a spectrum with peaks at $\omega_k$ of amplitude $a_k$. A regular orbit can be described by its three independent fundamental frequencies, meaning that all other peaks $\omega_k$ in the spectrum are a linear combination of those,  $\omega_k = \vec{n_k} \cdot \vec{\Omega}$, with $\vec{n}$ is a vector of three integers. If the fundamental frequencies are commensurable, i.e. one is a ratio of small integers of the other, $\vec{n} \cdot \vec{\Omega} = 0$, the orbit is said to be resonant. A singly-resonant orbit is confined to a two-dimensional surface, while a doubly-resonant orbit traces a closed one-dimensional curve. A sufficiently tight ensemble of particles can be resonantly trapped by a resonant orbit, meaning that their orbits have a finite libration amplitude around the resonance \citep{BinneyTremaine2008}. Depending on how strongly trapped the particles are, and thus how small the libration amplitude is, they will occupy a subspace of phase-space around the resonant orbit. While an ensemble of particles on regular non-resonant orbits phase-mixes at a rate~$\propto t^{-3}$, where $t$ is time, an ensemble of particles trapped by a resonant orbit phase-mixes slower, at a rate~$\propto t^{-2}$ for a singly-resonant orbit and at a rate~$\propto t^{-1}$ for a doubly-resonant orbit \citep{Vogelsberger2008}. 

\begin{figure*}[t!]
\centering
    \includegraphics[width=1\textwidth]{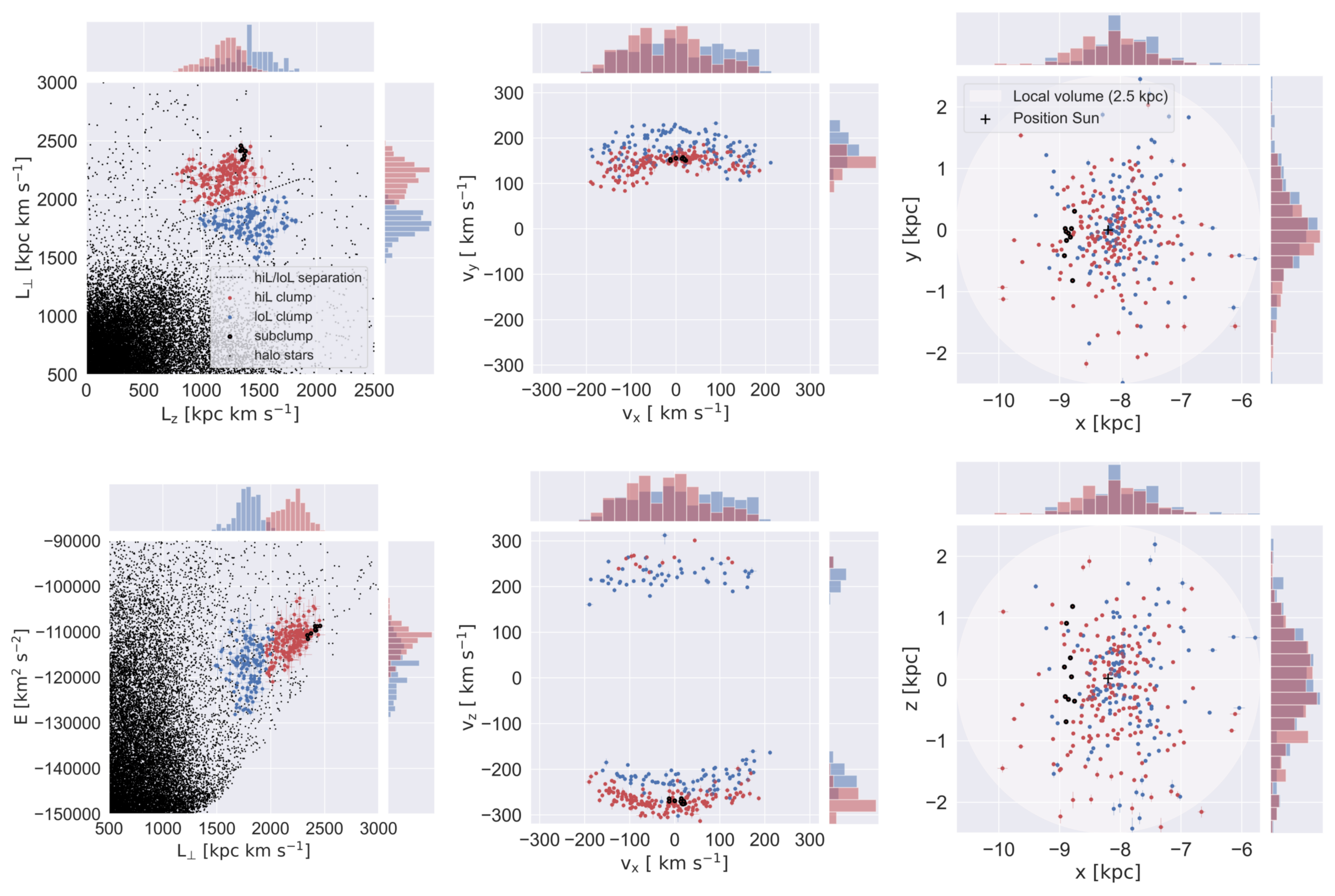} 
\caption{Distribution of the HS in IoM-space defined as $E, L_z$ and $L_{\bot}$ (left), velocity (middle) and configuration space (right) separated in the hiL clump (red, 191 stars in total) and loL clump (blue, 128 stars in total) based on the stars' position in $(L_z, L_{\bot})$ space, see top left panel. The subclump, a kinematically cold, coherent group of stars, is indicated in black. The stars in the background of the energy and angular momentum distributions correspond to the selection of local halo stars ($d$ < 2.5 kpc) by \cite{Dodd2023}. The energy values have been computed in the fiducial potential (see Appendix~\ref{app:pot}). Histograms on the top and right of each plot show the one-dimensional distributions. The error bars on the data points denote the uncertainties. The hiL clump has on average lower $L_z$, higher $L_{\bot}$, higher energy, larger $|v_z|$ and lower $v_y$. 
\label{fig:phasespace}}
\end{figure*}

At the edge of a family of resonant orbits we find a separatrix, which denotes a transition between orbit families. A separatrix is a stochastic region\footnote{Stochastic regions are found in near- or non-integrable potentials, and generally realistic galactic potential models are of this type.} hosting chaotic orbits. Such orbits are not quasi-periodic and do not have three IoM. Substructures on chaotic orbits therefore diffuse and mix faster than substructures on regular orbits \citep{PriceWhelan2016, Mestre2020}. Moreover, stellar streams on separatrices can exhibit unusual morphologies and similarly undergo faster diffusion \citep{Yavetz2021, Yavetz2022}. Naturally, triaxial potentials host a wider range of orbits, and consequently more separatrices and more chaotic orbits than an axisymmetric potential \citep{Papaphilippou1998, BinneyTremaine2008}. However, also axisymmetric potentials can host chaotic orbits \citep{Zotos2013, Zotos2014}, including simple models like the Miyamoto Nagai potential \citep{Pascale2022}. The location of chaotic regions, resonances, separatrices and different orbit families can be analysed using e.g. the orbital frequencies \citep{Valluri2012}, and these will of course depend on a potential's characteristics \citep{Caranicolas2010, Zotos2014}.

\subsection{Characterising the Helmi Streams}
\label{sec:HSprops}

Figure~\ref{fig:phasespace} shows the distribution of the HS stars in IoM ($E, L_z, L_{\bot}$), velocity and configuration space. The top left panel shows that the HS separate in two clumps in $(L_z, L_{\bot})$ space, as first identified by \cite{Dodd2022}. We empirically separate the sample into the hiL clump, having larger $L_{\bot}$, and loL clump, having lower $L_{\bot}$, and classify a star with a given $L_{\bot, *}$ and $L_{z, *}$ as follows
\begin{equation}
L_{\bot, *} \left\{
    \begin{array}{ll}
        > 0.35 \cdot L_{z, *} + 1525~\si{\:kpc \: km \: s^{-1}} & \rightarrow \texttt{hiL clump} \\
        < 0.35 \cdot L_{z, *} + 1525~\si{\:kpc \: km \: s^{-1}} & \rightarrow \texttt{loL clump},
    \end{array}
\right.
\label{eq:HShiLloL}
\end{equation}
\noindent see also Fig.~\ref{fig:phasespace}. There are 191 and 128 stars in the hiL and loL clumps (respectively)\footnote{The reason that there are more hiL stars than loL stars, a ratio of $\sim 2:3$, could be caused by the fact that the loL clump is located closer to the disc in integrals of motion space, which results in a stronger background, possibly making it harder for the clustering algorithm to pick up the structure.}. The median position of the hiL stars is $(L_z, L_{\bot}) = (1234, 2207) \si{\: kpc \: km \: s^{-1}}$, while the median position of the loL stars is  $(L_z, L_{\bot}) = (1436, 1791) \si{\: kpc \: km \: s^{-1}}$. We confirm that these two clumps have consistent stellar populations and metallicity distributions, with a median metallicity of -1.45~dex and standard deviation of 0.42~dex using LAMOST LRS DR7 metallicity estimates \citep{zhao2012}.

\begin{figure}[t!]
\centering
    \includegraphics[width=0.4\textwidth]{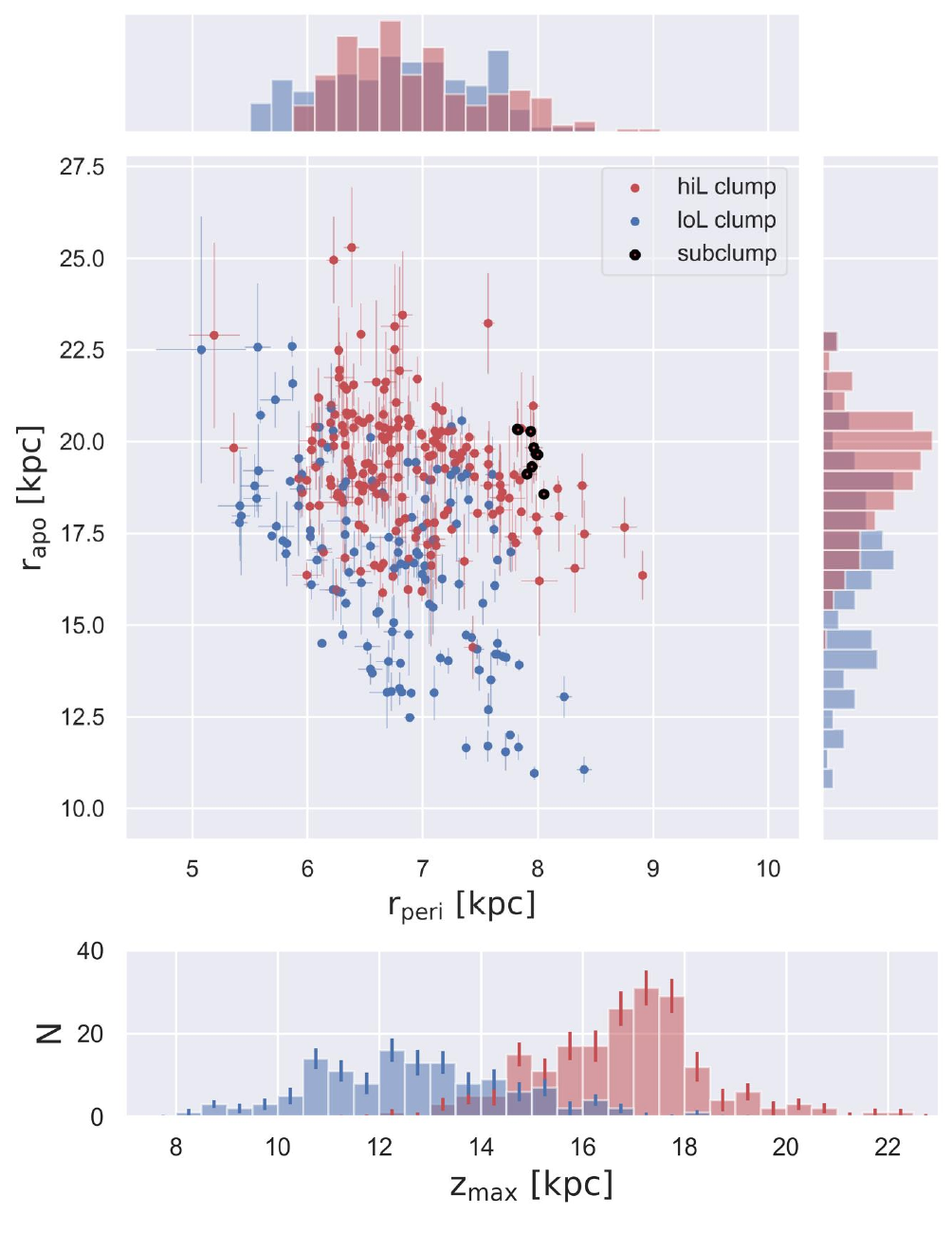} 
\caption{Pericentre, apocentre and $z_{max}$ distribution of the HS' orbits in the fiducial potential (see Appendix~\ref{app:pot}). The hiL stars are indicated in red, the loL stars in blue and the subclump with black edges as in Fig.~\ref{fig:phasespace}. Histograms on the top and right of the top plot show the one-dimensional distributions. The apocentres and pericentres were determined as the maximum and minimum distance from the Galactic Centre that the orbit reached over an integration time of 10 Gyr. An estimate for the uncertainties was obtained by randomly sampling the observables a 1000 times within their uncertainties after integrating their orbits.
\label{fig:orbparams}}
\end{figure}

The lower left panel of Fig.~\ref{fig:phasespace}, displaying the ($E, L_{\bot})$ distribution, shows that on average the hiL stars are less bound and the loL stars span a larger range in energy. The hiL stars have larger apocentres and reach higher above the midplane, see also Fig.~\ref{fig:orbparams}. This is robust for different Galactic potential models, though of course the exact apo- and pericentre distribution changes.  

The two middle panels of Fig.~\ref{fig:phasespace} show the velocity distribution of the two clumps. Interestingly, the ratio of stars with positive $v_z$, to  negative $v_z$ differs between the two clumps. For the hiL clump this ratio is $\dfrac{N_{\rm{hiL}}(v_{z}^{+})}{N_{\rm{hiL}}(v_{z}^{-})} = \dfrac{13}{178} \sim 0.07$, while for the loL clump this ratio is $\dfrac{N_{\rm{loL}}(v_{z}^{+})}{N_{\rm{loL}}(v_{z}^{-})}= \dfrac{49}{79} \sim 0.62$. In earlier work, the bimodality in $v_z$ of the entire HS has been used to obtain a rough estimate for the time of accretion \citep{Kepley2007, Koppelman2019Helmi}, since it is an indication of a structure's degree of phase-mixing. The hiL clump not only appears to be less phase-mixed, but it forms a tighter, more coherent structure in velocity space than the loL clump stars. The hiL clump even exhibits a kinematically cold subclump (in black in Fig.~\ref{fig:phasespace} and Fig.~\ref{fig:orbparams}), and defines a stream-like structure in configuration space, as the two right panels of Fig.~\ref{fig:phasespace} show. 

We select the subclump as $1320 \leq L_z \leq 1410 \si{\: kpc \: km \: s^{-1}}$, $2330 \leq L_{\bot} \leq 2500 \si{\: kpc \: km \: s^{-1}}$ and $x > - 9.2$~kpc and find 9 members. Four of these stars have LAMOST LRS DR7 metallicity estimates \citep{zhao2012}, which are [Fe/H] $= -1.77$, $-2.04$, $-1.48$ and $-2.19$ dex, with typical uncertainties of 0.1~dex \citep{Anguiano2018, Niu2023} to $\sim 0.3$~dex at the metal-poor end \citep{Li2018VMP}. Given these uncertainties and the metallicity distribution of the HS, the subclump is indistinguishable from the HS in its stellar populations. Although this may apparently contrast \cite{Myeong2018}' \enquote{independent} identification of the subclump in \textit{Gaia} DR2, naming it S2, subsequent high-resolution spectroscopic follow-up of S2 by \cite{Aguado2021} also support the association. This has been convincingly demonstrated by \cite{Matsuno2022} who studied the abundance patterns of a sample of HS stars and homogeneously reanalysed the spectra of three \cite{Aguado2021} stars and concluded that their abundance patterns are fully consistent.  Similarly, \cite{Dodd2022} identified and considers the subclump as a part of the HS. Given its much tighter distribution, the subclump must have evolved dynamically more slowly than the rest of the substructure, as it is significantly less phase-mixed. This could happen if the dynamical evolution of the subclump is affected by a resonance.

\subsection{Orbital characterisation: frequency analysis and chaos indicators}

In cylindrical coordinates, we typically use three orbital frequencies ($\Omega_R, \Omega_{\phi},\Omega_{z}$) which are related to the orbit's oscillation in the radial, vertical and azimuthal directions. For a regular orbit, the components of its frequency spectrum are linear integer combinations of the three orbital frequencies $(\Omega_R, \Omega_{\phi}, \Omega_z)$\footnote{This set of three orbital frequencies is not necessarily equal to the three fundamental orbital frequencies discussed in Sect. \ref{sec:review}.}. For a resonant orbit, two or more frequencies are commensurate, while for a chaotic orbit the orbital frequencies vary with time. To determine the orbital frequencies corresponding to a star's orbit, we use a modified version of \texttt{SuperFreq} \citep{SuperFreq}, as described in detail in Appendix~\ref{app:freqsmethod}.

In contrast to regular orbits, a chaotic orbit 
has less than 3 integrals of motion, and as a consequence it undergoes orbital diffusion. Therefore, if the MW's potential hosts a stochastic region in between the two HS clumps, this could possibly lead to a depletion of stars in that region. To identify chaotic behaviour, we make use of the Lyapunov exponent~$\Lambda$. Take an orbit $\vec{x}(t)$ and an orbit infinitely close to it, $\vec{x}(t) + \vec{w}(t)$. Here, $\vec{w}(t)$ is the so-called deviation vector, which grows as a power-law in time for a regular orbit, but exponential in time for a chaotic orbit. The Lyapunov exponent $\Lambda$ measures the time variation of $\vec{w}(t)$,
\begin{equation}
    \Lambda \equiv \lim_{t\to\infty} \dfrac{\ln | \vec{w}|}{t}
\end{equation}
\noindent To compute $\Lambda$, we resort to its finite-time estimate. For a period of time where the orbit is regular, $\ln | \vec{w}| / t$ fluctuates around a constant  value, and in that case $\Lambda$ is set to zero. When an orbit is chaotic, $\Lambda$ is estimated as the average value of  $\ln | \vec{w}| / t$ over the period of exponential growth. Thus, the larger $\Lambda$ is, the more chaotic the orbit is. The Lyapunov exponent also gives an indication of the timescale of chaoticity. In this work, the finite-time estimate of $\Lambda$ is computed using \texttt{AGAMA} over an integration time of 100~Gyr.

\begin{figure*}
\centering
    \includegraphics[width=0.9 \hsize]{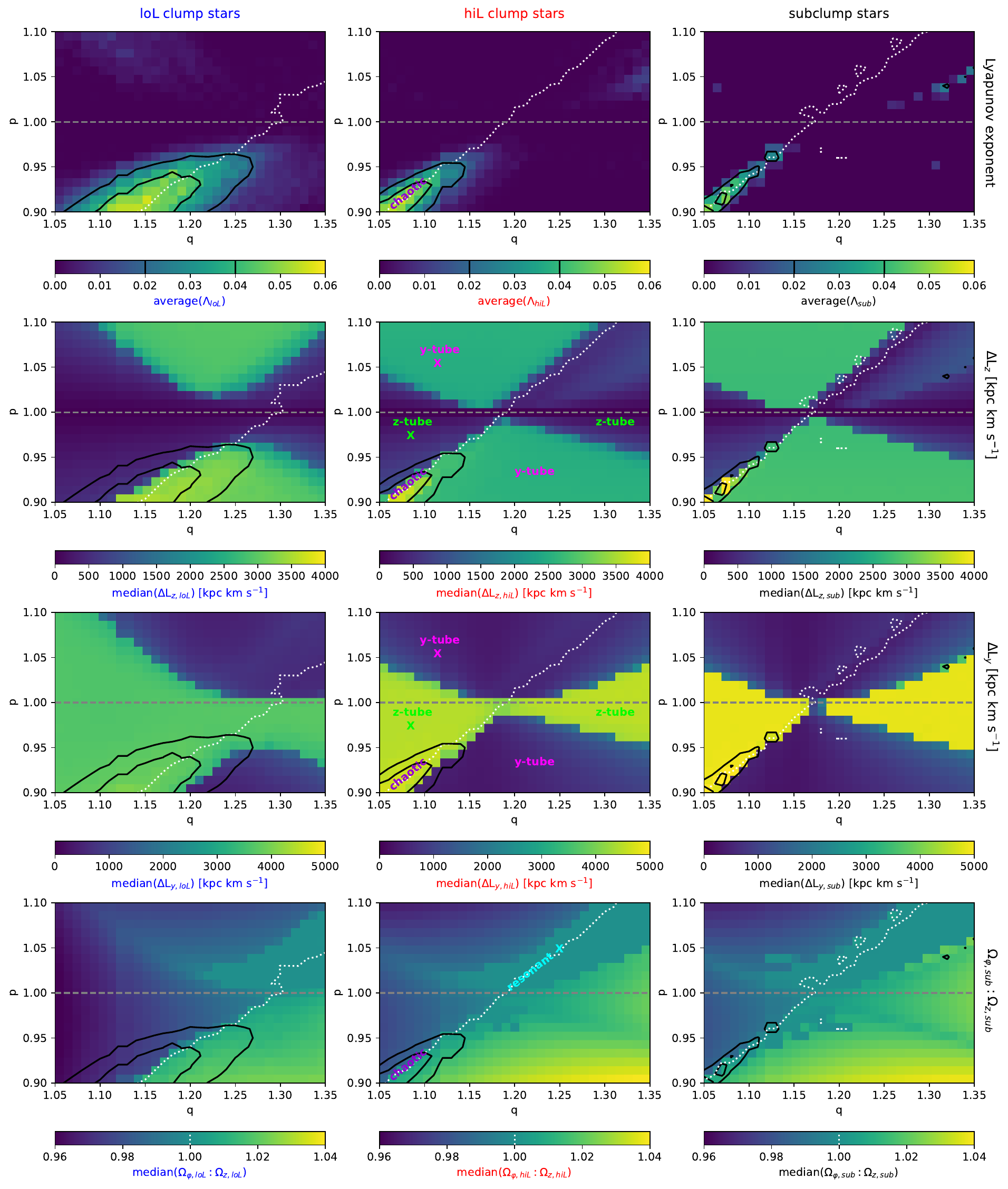}
\caption{Various dynamical quantities of the loL stars (left column) and hiL stars (middle column) and subclump stars (right column) in a potential with a triaxial halo with a range of $p$ and $q$ as calculated by \texttt{AGAMA} for an integration time of 100~Gyr. The grey horizontal line at $p = 1$ serves to guide the eye and indicates axisymmetric potentials. The white dotted line indicates the $\Omega_{\phi} : \Omega_z$ = 1:1 resonance (see also bottom row). \textit{Top row}:  average Lyapunov exponent. A non-zero Lyapunov exponent indicates that there is chaoticity. The overlaid black contours, which are also shown in the other rows, map \texttt{average}($\Lambda_*$) at the levels of 0.02 and 0.04, as is also indicated on the colourbar. \textit{Second row:} median variation in $L_z$, with $\Delta L_{z, *} = \max (L_{z,*}) - \min (L_{z, *})  \:$ over the entire integration time. \textit{Third row:} median variation in $L_y$, with $\Delta L_{y, *} = \max (L_{y,*}) - \min (L_{y, *})  \:$  over the entire integration time. \textit{Bottom row:} \texttt{median($\Omega_{\phi}:\Omega_z$)} ratio. We can see clear correspondences between the four different rows. The chaotic regions overlap with the \texttt{median($\Omega_{\phi}:\Omega_z$)} = 1:1 resonance and a transition in the value of $\Delta L_z$ and $\Delta L_y$. This transition indicates a change in orbit family, and the text in the middle column indicates where the different orbit families reside, of which examples are shown in Fig.~\ref{fig:bigorbfig} (the crosses in this figure indicate their $p$ and $q$). To the left of the chaotic band we find $z$-tube orbits, which have an oscillating $L_y$ and a roughly conserved $L_z$, such that \texttt{median($\Delta L_y$)}  $\lesssim 4000 \si{\: kpc \: km \: s^{-1}}$, \texttt{median($\Delta L_z$)}  $\lesssim 500 \si{\: kpc \: km \: s^{-1}}$. To the right of the chaotic band we find $y$-tube orbits, which have a roughly conserved $L_y$ and an oscillating $L_z$, such that \texttt{median($\Delta L_y$)}  $\sim 1000 \si{\: kpc \: km \: s^{-1}}$ and \texttt{median($\Delta L_z$)}  $\sim 3000 \si{\: kpc \: km \: s^{-1}}$. For $p > 1$, the $\Omega_{\phi} : \Omega_z$ = 1:1 is a resonance that traps orbits, which reduces their variation in both $L_y$ and $L_z$. This is most clear for the subclump stars.
    \label{fig:mapstogether}}
\end{figure*}
\section{Analysis} \label{sec:analysis}

\subsection{The HS' orbits in a triaxial potential}\label{sec:chaos}

To investigate how the dynamical properties of the HS can be explained, we study their orbits in a range of mildly triaxial potentials, and we explore the orbital structure in the region of phase space occupied by the HS. We explore triaxial potentials as,  on the basis of cosmological simulations, it is likely that DM halos have a triaxial shape, particularly at larger radii \citep{Frenk1988, VeraCiro2011DMhalo, Cataldi2021}. Moreover, there is evidence that Milky Way's halo is being perturbed and deformed by the infall of the Large Magellanic Cloud \citep[LMC,][]{GaravitoCamargo2019, GaravitoCamargo2021}.

We modify the fiducial MW model based on the \cite{McMillan2017} potential described in Sect.~\ref{sec:datamethod:generalities} (see also Appendix~\ref{app:pot}) by replacing its NFW halo with a triaxial halo:

\begin{equation}
    \rho_{\rm{NFW}}(x, y, z) = \frac{\rho_0}{\dfrac{\tilde{r}}{r_s} \left( 1 + \dfrac{\tilde{r}}{r_s}   \right)^2}, 
\label{eq:NFW}
\end{equation}
\noindent with
\begin{equation}
    \tilde{r} = \sqrt{ x^2 + \dfrac{y^2}{p^2} + \dfrac{z^2}{q^2} }
\end{equation}
\noindent and where
\begin{equation}
    p^2 = \dfrac{b^2}{a^2}, \: \: \: q^2 = \dfrac{c^2}{a^2} \: \: \: \text{.}
\end{equation}
\noindent We make sure that the circular velocity at the position of the Sun is equal to $v_{\rm{c}}({R_\odot}) = 233.2 \si{\:km \: s^{-1}}$ when varying ($p, q$)\footnote{To this end, we update the initial density $\rho_{\rm{0,i}}$ of a DM halo with a given  $p$, $q$ to $\rho_{\rm{0, new}}$, while we keep the other components of the potential fixed. We use that $v_{\rm{c}}(R_{\odot})^2 = v_{\rm{c, baryons}}(R_{\odot})^2 + v_{\rm{c, halo}}(R_{\odot})^2$, where $v_{\rm{c, baryons}}(R_{\odot})$ is the contribution of all baryonic components to $v_{\rm{c}}(R_{\odot})$, which is fixed, and $v_{\rm{c, halo}}(R_{\odot})$ is the contribution of the DM halo. Since $v_{\rm{c, halo}}(R_{\odot})$ follows the proportionality $v_{\rm{c, halo}}(R_{\odot})^2 \propto \rho_0$, we compute the density $\rho_{\rm{0, new}} = \dfrac{233.2^2 - v_{\rm{c, baryons}}(R_{\odot})^2}{v_{\rm{c, halo}}^2({R_\odot})}  \rho_{\rm{0,i}}$.}. We explore the range $1.05 \leq q \leq 1.35$ and $0.90 \leq p \leq 1.10$, which is motivated by disc stability constraints \citep{Prada2019, Cataldi2021} and the requirement that to maintain the gap we need an effective potential that is roughly spherical \citep{Dodd2022}. To study how varying $p$ and $q$ influences HS' orbits, we integrate these for 100~Gyr and compute (1) the finite-time estimate of the  Lyapunov exponent to study chaoticity, (2) we quantify the variation in $L_z$ and $L_y$ for each orbit to distinguish different types of orbit families, and (3) we determine the orbital frequencies $\Omega_{R}$, $\Omega_{\phi}$ and $\Omega_{z}$ as outlined in Appendix~\ref{app:freqsmethod} to identify possible orbital resonances. The average/median values of these quantities for the loL stars, hiL stars and subclump stars are shown in the left, central and right columns of Fig.~\ref{fig:mapstogether} respectively. We discuss the behaviour of each dynamical quantity in the following paragraphs.

\subsubsection{Chaoticity: the Lyapunov exponent}

The average Lyapunov exponent of the hiL stars, loL stars and subclump are shown separately in the form of a map in the top row of Fig.~\ref{fig:mapstogether}. There is a band of chaoticity (with light green-yellow corresponding to $\Lambda \gtrsim 0.04$) in the region $q \sim 1.13$, $p \sim 0.90$, till $q \sim 1.25$, $p \sim 0.97$ for the loL stars, while for hiL stars the chaoticity band occupies a  narrower region from $q \sim 1.05$, $p \sim 0.90$  till  $q \sim 1.15$, $p \sim 0.97$, and an even narrower region with chaotic behaviour is present for $q \sim 1.05$, $p \sim 0.90$,  till  $q \sim 1.13$, $p \sim 0.97$ for the subclump stars. The band of chaoticity is thus located at larger $q$ for the loL stars, and it also appears broader. This is due to the loL stars occupying a larger (and different) volume in phase space than the hiL stars, see Fig.~\ref{fig:phasespace}. Interestingly, there is no analogous chaotic region for potentials with $p > 1$.

\subsubsection{$y$-tube and $z$-tube orbits and variations in $L_z$ and $L_y$}

\begin{figure*}[!t]
\centering
    \includegraphics[width=0.95\hsize]{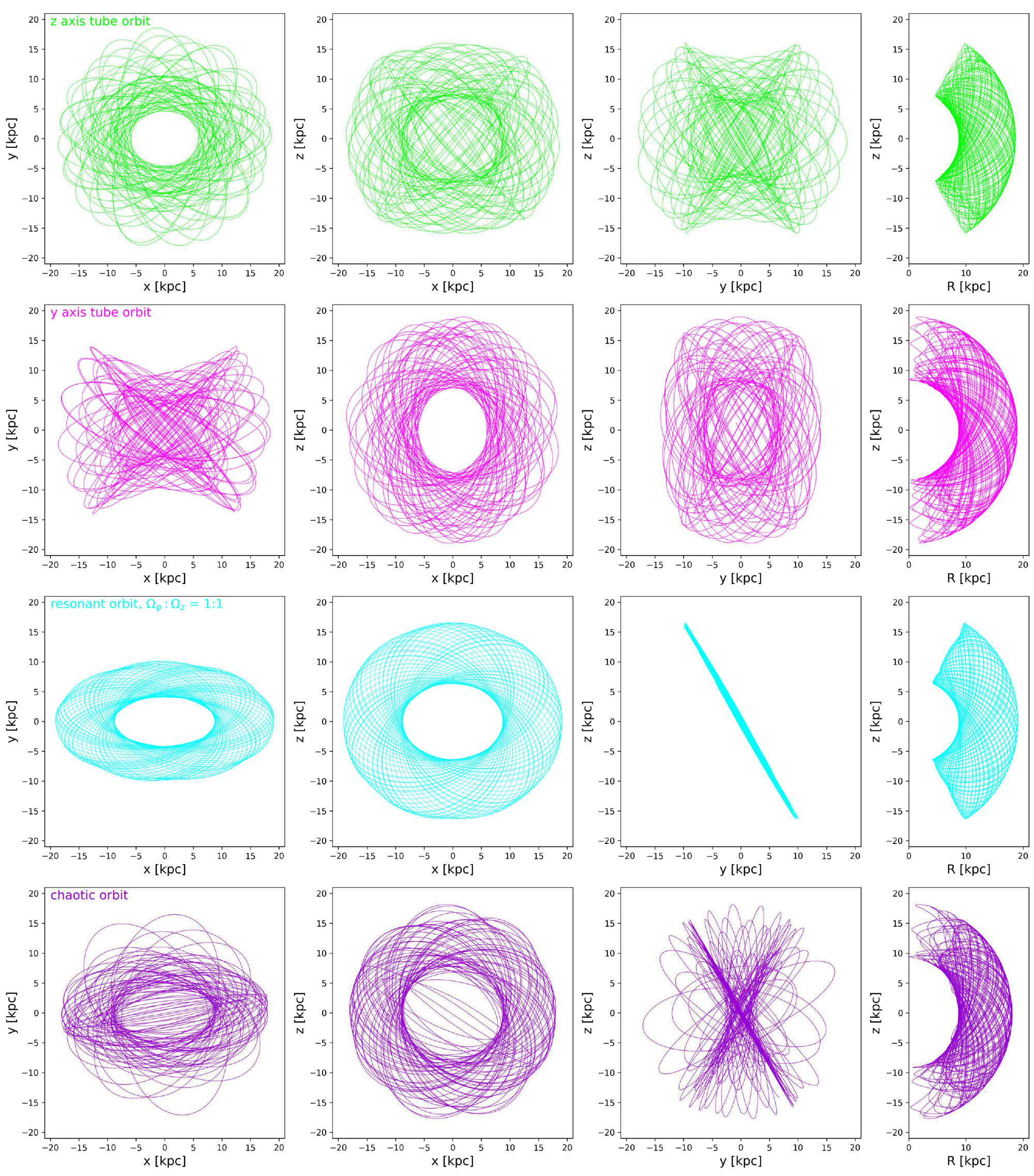}
\caption{Orbits integrated for 25~Gyr for a randomly selected hiL star in different triaxial potentials: {\it top row:} $p=0.97, q=1.08$, resulting in a $z$-tube orbit( green); {\it second row:} a $y$-tube orbit for $p=1.05, q = 1.11$ (pink); {\it third row:} a resonant orbit for $p=1.05, q=1.245$ (cyan); {\it bottom row:} a chaotic orbit with $\Lambda = 0.12$ for $p=0.83, q = 0.977888$. 
\label{fig:bigorbfig}}
\end{figure*}

We can get more insight into what causes the band of chaoticity by inspecting the orbits of the HS stars in the different potentials, of which examples are shown in Fig.~\ref{fig:bigorbfig} (see also Appendix~\ref{appendix:maps}). The tube orbit family comprises regular orbits which fill a doughnut-like shaped volume in configuration space. Tube orbits have a sense of rotation around the principal axes of the potential, and depending on which axis they are called short- or inner/outer long axis tube orbits \citep[see also][]{BinneyTremaine2008}. As we vary $p$ and $q$, the long and short axis swap orientation\footnote{For example, in a simple axisymmetric case, $z$ is the short axis if $q < 1$, while $z$ is the long axis if $q > 1$.}. Therefore, we choose a more general naming convention and define the $z$-axis tube orbits ($z$-tube orbits) as rotating around the $z$-axis, while the $y$-axis tube orbits ($y$-tube orbits) are defined as rotating around the $y$-axis. Tube orbits are centrophobic, i.e. they do not pass through the centre of the potential, which results in a non-zero time-averaged angular momentum about whichever axis they rotate. This can also be appreciated from Fig.~\ref{fig:bigorbfig} and \ref{fig:bigorbfig_LzLperp}, which shows that the $z$-tube orbit rotates around the $z$-axis, keeping its $L_z$ roughly conserved in time, while its $L_y$ varies by a large amount with a time-average equal to zero. On the other hand, the $y$-tube orbit revolves around the $y$-axis, keeping its $L_y$ roughly conserved, while its $L_z$ varies by a large amount with a time-average equal to zero.

A way to differentiate between these two orbit families is thus by inspecting the orbit's variation in $L_z$ and $L_y$, which we define as
\begin{equation}
    \Delta L_{i} = \max (L_{i}) - \min (L_{i})
    \label{eq:deltaL}
\end{equation}
\noindent where $i$ refers to the component of the angular momentum, in this case $y$ or $z$, and $\Delta L_i$ is calculated over at least one period in angular momentum space (e.g. the period of the $z$-tube orbit shown in Fig.~\ref{fig:bigorbfig_LzLperp} is about 25~Gyr). Figure~\ref{fig:bigorbfig_LzLperp} illustrates the expected range of variation in $L_y$ and $L_z$: the $z$-tube orbit has $\Delta L_z \approx 500 \si{\: kpc \: km \: s^{-1}}$ and $\Delta L_y \approx 4500 \si{\: kpc \: km \: s^{-1}}$, while the $y$-tube orbit has $\Delta L_z \approx 3000 \si{\: kpc \: km \: s^{-1}}$ and $\Delta L_y \approx 500 \si{\: kpc \: km \: s^{-1}}$. The bottom row of Fig.~\ref{fig:bigorbfig} shows a chaotic orbit without a third IoM which has $\Delta L_z \approx \Delta L_y \approx 4500 \si{\: kpc \: km \: s^{-1}}$.

The second row of Fig \ref{fig:mapstogether} shows a map of the \texttt{median}($\Delta L_z$) for the hiL stars and loL stars for different $p$ and $q$, while the third row of Fig.~\ref{fig:mapstogether} shows the \texttt{median}($\Delta L_y$). These maps clearly show the regions occupied by the $z$-tube and $y$-tube orbits. A comparison of the top three rows of Fig.~\ref{fig:mapstogether} reveals that the chaotic band coincides with the transition between these two orbit families. In this chaotic band, \texttt{median}($\Delta L_z$) is larger, as expected. The maps also show that potentials with DM halo shapes $p >1 $ and $p <1 $ have a roughly similar behaviour and largely mirror each other, but for $p > 1$ there is a region with a lower \texttt{median}($\Delta L_z$) that is not present in $p < 1$. This region corresponds to an orbital resonance that traps orbits, as we discuss in the next section.

Summarising, Fig.~\ref{fig:mapstogether} shows that there is a part of $(p, q)$ parameter space where the hiL stars are on $y$-tube orbits, while the loL stars are on $z$-tube orbits. This holds roughly from $(p, q) = (0.9, 1.10)$ till $(p, q)= (0.98, 1.17)$ and from $(p, q) = (1.01, 1.20)$ till $(p, q)=(1.10, 1.05)$. In such a configuration, we would expect that the hiL clump stars slowly change their $L_z$ over time, while the loL clump's $L_z$ stays roughly conserved, leading to a separation of the two clumps over time. Such behaviour could therefore  possibly explain the formation of two (the hiL and loL)  clumps.

\subsubsection{Orbital frequencies and resonances}
\label{sec:freqsres}

To investigate further the transition in orbit family at these specific $p$ and $q$, we inspect the behaviour of the orbital frequencies $\Omega_R$ , $\Omega_{\phi}$, and $\Omega_z$. The bottom row of Fig.~\ref{fig:mapstogether} shows a map of the ratio of $\Omega_{\phi}:\Omega_z$ for the values of $p$ and $q$ explored in the other panels. This reveals that the change in orbit family seen in the two middle rows is related to the $\Omega_{\phi} : \Omega_z$ = 1:1 resonance. For $p < 1$, the chaotic band neatly overlaps with the white dotted
line indicating the resonance. Therefore, the chaotic behaviour seen in the top row is due to the stochastic layer around the 1:1 resonance. This happens for $p < 1$ because the short and long axis of the potential are exchanged: for an oblate effective potential, $z$ is the short axis, while for $p <1$ and a prolate effective potential, the short axis is $y$. Because of this, the 1:1 resonance works as a separatrix: stars on this resonance do not know whether to rotate around the $z$-axis or $y$-axis, as the chaotic orbit in Fig.~\ref{fig:bigorbfig} shows. This orbit seems to jump between a $y$-tube orbit (0-10~Gyr), $z$-tube orbit (10-20~Gyr) and a resonantly trapped orbit (20-25~Gyr).

\begin{figure}[t!]
\centering
    \includegraphics[width=\hsize]{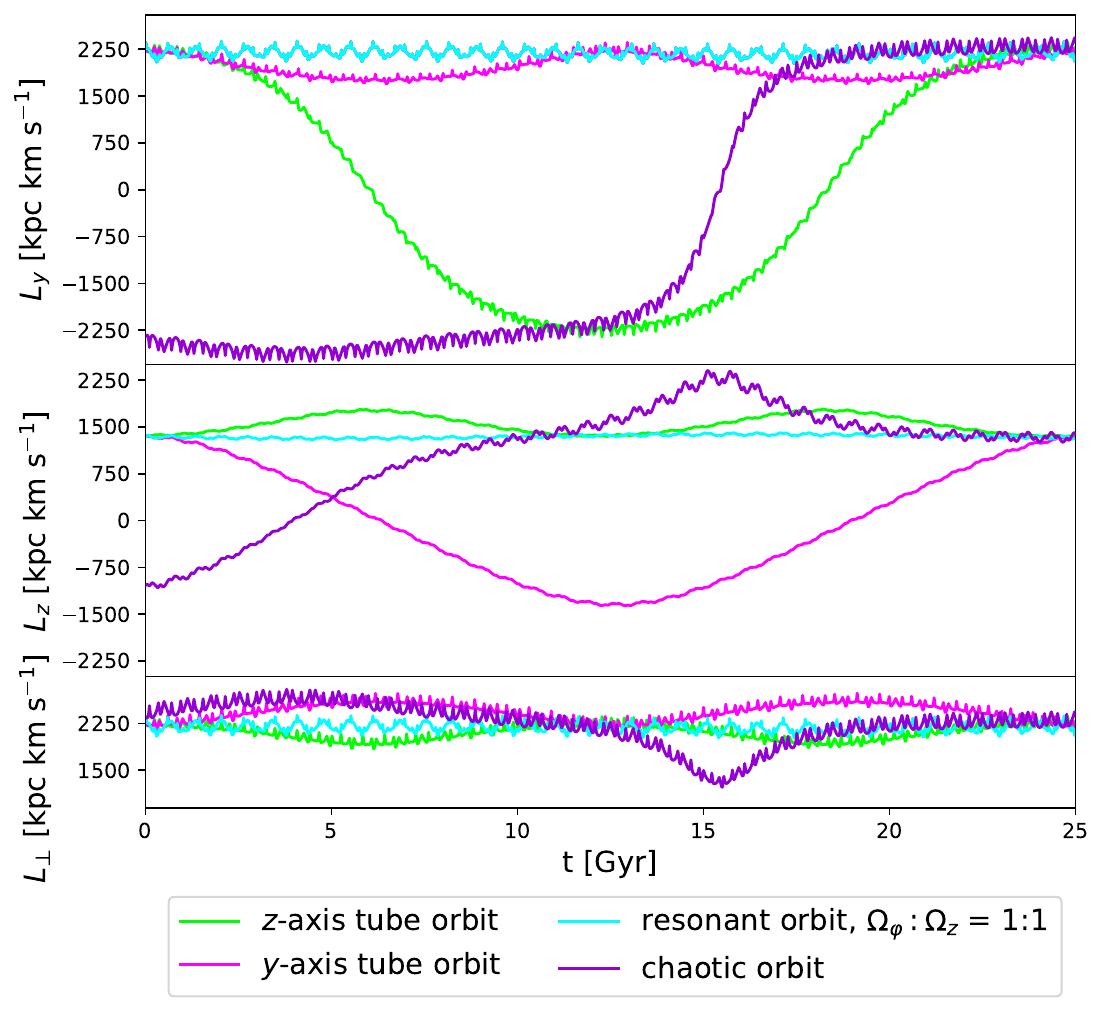}
\caption{Behaviour over time of $L_z$, $L_y$ and $L_{\bot}$ for the orbits shown in Fig.~\ref{fig:bigorbfig}. Since the $z$-tube orbits around the $z$-axis, it has a roughly conserved $L_z$ and $L_{\bot}$ but a time-varying $L_y$. Similarly, since the $y$-tube orbits around the $y$-axis, it has a roughly conserved $L_y$ and $L_{\bot}$ but a time-varying $L_z$. For the chaotic orbit, $L_z$, $L_y$ and $L_{\bot}$ vary largely in time. The resonant orbit has a small variation in $L_y$, $L_z$ and $L_{\bot}$.
\label{fig:bigorbfig_LzLperp}}
\end{figure}

On the other hand, for $p >1$, the region that has a lower \texttt{median}($\Delta L_z$) overlaps with the 1:1 resonance, and in this case the resonance actually stabilises the orbits and resonantly traps them. These resonantly trapped orbits have angular momenta both around the $z$ and $y$-directions, and orbit in a plane in $(y, z)$, see also Fig.~\ref{fig:bigorbfig} and Appendix~\ref{appendix:maps}. Their $L_y, L_z$ and $L_{\bot}$ are almost constant in time, as seen in Fig.~\ref{fig:bigorbfig_LzLperp}, indicating that in this region the effective potential is roughly spherical.

Given these findings, a picture emerges where the loL and hiL clump could be on different orbital families. This would require a potential with ($p, q$) values such that the hiL clump is on the $y$-tube orbits and the loL clump on the $z$-tube orbits. Furthermore, the presence of the kinematically cold subclump in the hiL clump suggests that it is associated with a stabilising orbital resonance, as present for $p >1$, as this will slow down phase mixing. From these considerations, it naturally follows that the subclump could be on the $\Omega_{\phi} : \Omega_z$ = 1:1 resonance, while (part of) the hiL clump would be resonantly trapped by that same resonance. The loL clump would not be on the 1:1 resonance, and would phase-mix at a normal rate as this would explain the asymmetries reported in Sec.~\ref{sec:HSprops} regarding the number of stars in the streams with positive and negative $z$-velocities. The opposite effect would occur for $p<1$, as the orbits are chaotic close to the $\Omega_{\phi} : \Omega_z$ = 1:1 resonance and undergo quick orbital diffusion, and a kinematically cold subclump could not be sustained. This is why $p < 1$ is disfavoured.

\begin{figure*}[t!]
\centering
    \includegraphics[width=\hsize]{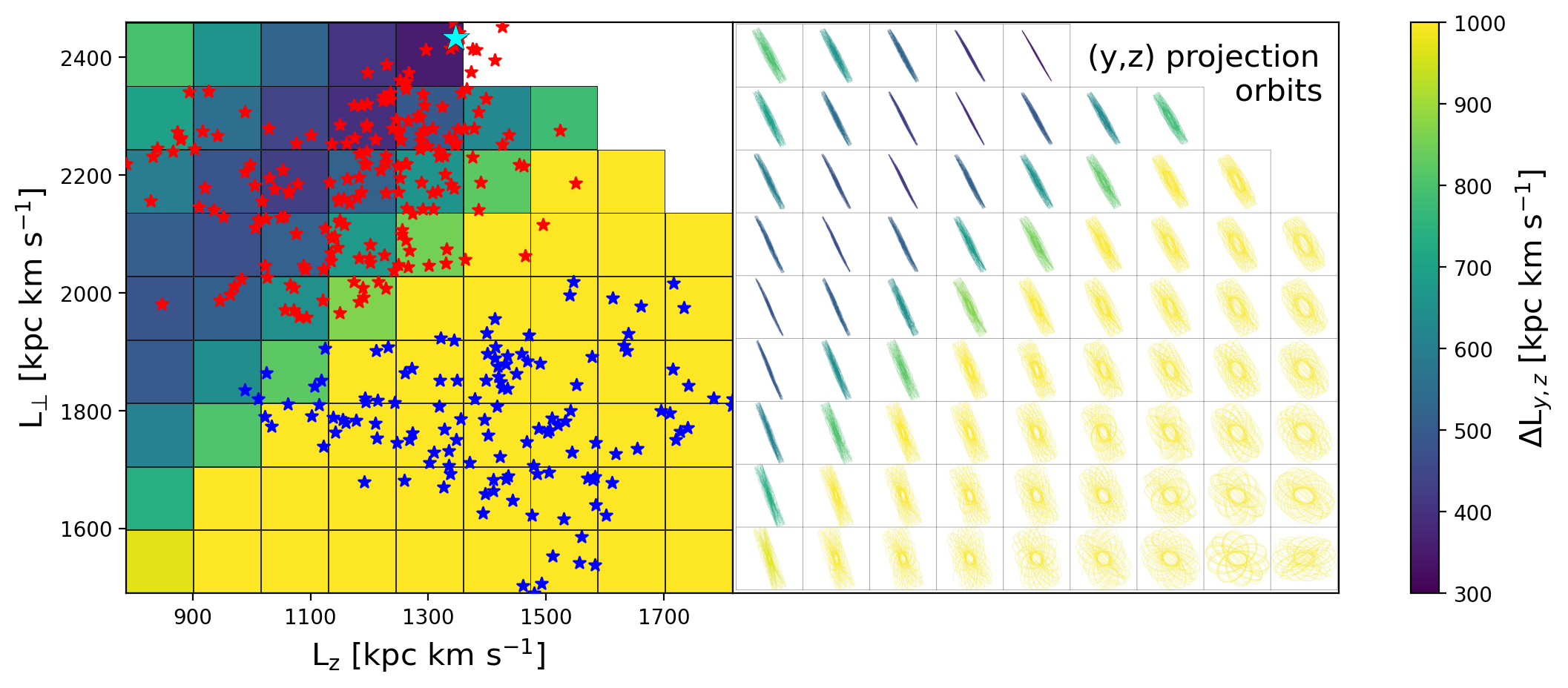}
\caption{The left panel depicts with different colours the variation in angular momentum, $\Delta L_{y, z} = \sqrt{\Delta L_{y}^2 + \Delta L_{z}^2}$, at different $L_z$ and $L_{\bot}$ but at a fixed energy and position in a potential with a triaxial NFW halo with $p = 1.02$, $q = 1.19$.  The 6D phase space coordinates of a subclump star, which is resonantly trapped in this potential and shown with the cyan star symbol, were used to set the energy and position of the orbits. $\Delta L_{y, z}$ is a proxy for how resonantly trapped an orbit is by the $\Omega_{\phi} : \Omega_z$ = 1:1 resonance, with a lower value indicating a more resonantly trapped orbit. For reference, the hiL stars and loL stars are shown as red and blue star symbols, respectively. The right panel shows orbits in Galactocentric Cartesian coordinates ($y, z$) corresponding to the fixed energy, position and different $L_z$ and $L_{\bot}$ as given by the grid on the left panel. Orbits that are on the $\Omega_{\phi} : \Omega_z$ = 1:1 resonance appear as a line in this plane, as they are two-dimensional structures. Resonantly trapped orbits occupy a larger volume and are flattened three-dimensional structures. 
\label{fig:sizeresonance}}
\end{figure*}

\subsubsection{Resonance trapping}
\label{sec:restrap}

We study the effect of resonance trapping by the $\Omega_{\phi} : \Omega_z$ = 1:1 resonance in more detail in Fig.~\ref{fig:sizeresonance}. We integrate a set of orbits in a potential with a triaxial halo with $p = 1.02$, $q = 1.19$, as in this potential the subclump stars are strongly resonantly trapped (see for example Appendix~\ref{appendix:maps}). The orbits' initial conditions are generated by varying ($L_z, L_{\bot}$) on a grid and using the energy and 3D position of a subclump star. We inspect the variation in angular momentum,  $\Delta L_{y, z} = \sqrt{\Delta L_{y}^2 + \Delta L_{z}^2}$, to quantify how strongly trapped an orbit is. Orbits on the $\Omega_{\phi}~:~\Omega_z$ = 1:1 resonance have a roughly conserved $L_{y}$ and $L_{z}$, while $z$-tube orbits and $y$-tube orbits do not conserve their $L_y$ and $L_z$, respectively (see Fig.~\ref{fig:bigorbfig_LzLperp}). Hence, a small value of $\Delta L_{y, z}$ indicates that an orbit is resonantly trapped, allowing us to probe the extent of the resonance, as we show in Fig.~\ref{fig:sizeresonance}. 

Figure~\ref{fig:sizeresonance} shows that the resonance covers a region that is as large as the hiL clump, though not each orbit is as strongly trapped as the other. The smaller the variation in angular momentum $\Delta L_{y, z}$ is, the more strongly resonantly trapped the orbit is, as can be seen by comparing the left and right panels of Fig.~\ref{fig:sizeresonance}. The right panel shows the $(y, z)$ projection of the orbits over an integration time of 20~Gyr, and reveals that resonant orbits define a 2D planar structure in $(y, z)$ space. 

\begin{figure*}[t!]
\centering
    \includegraphics[width=\hsize]{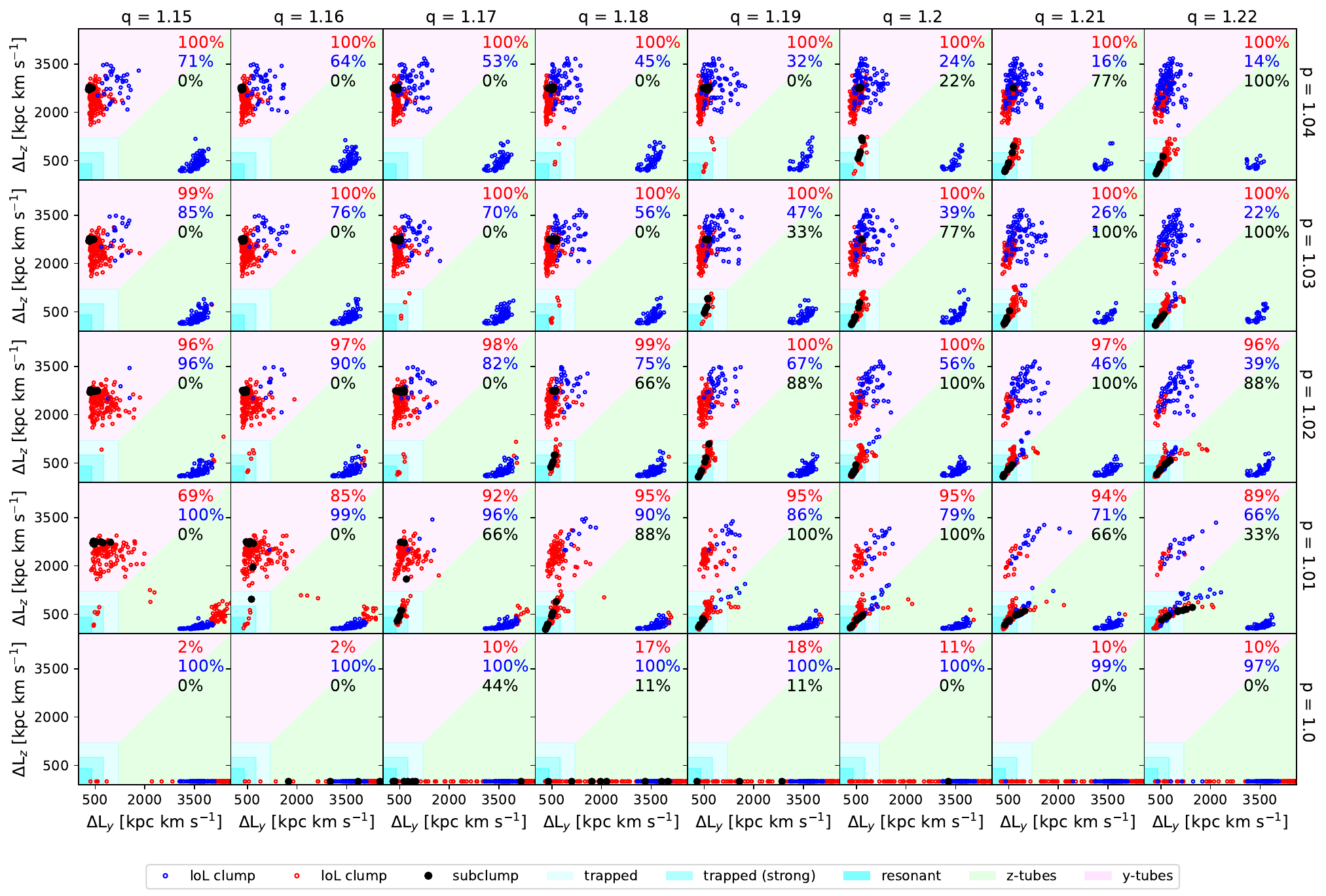}
\caption{$\Delta L_y$ and $\Delta  L_z$ calculated over an integration time of 100~Gyr for the orbits  of the HS' stars in a potential with a triaxial NFW halo for a range of $1.00 < p < 1.04$ and $1.13 < q < 1.23$ . The hiL stars are plotted in red, the loL stars in blue and the subclump stars in black. The percentages indicate how many percent of the hiL stars are on $y$-tube orbits or resonantly trapped orbits (red percentage), how many loL stars are on $z$-tube orbits (blue percentage) and how many subclump stars are strongly resonantly trapped (black percentage). The coloured backgrounds indicate regions occupied by different orbit families (see Eq. \ref{eq:orbs}): resonantly trapped orbits, strongly resonantly trapped orbits and resonant orbits, in light blue, blue and darker blue, $y$-tube orbits in pink and $z$-tube orbits in green.
\label{fig:HS_LzLyvar}}
\end{figure*}

\subsubsection{The hiL and loL clump on different orbit families}
\label{sec:analysis:differentorbfams}

To understand in which potential the individual hiL and loL clump stars are on the $y$-tube, $z$-tube or resonantly trapped orbits, we compute $\Delta L_z$ and $\Delta L_y$ for each individual HS star in the potentials for the range  $1.00 < p < 1.04$ and $1.13 < q < 1.23$. Based on visual inspection of a large range of orbits, we employ the following empirical criteria based on the orbit's $\Delta L_z$ and $\Delta L_y$ (see Fig.~\ref{fig:bigorbfig} and Sect.~\ref{sec:chaos}) to call them:
\begin{equation}
\begin{array}{l}
\texttt{orbit} \\ \texttt{family}
\end{array}
= \left\{
    \begin{array}{l}
        \texttt{z-tubes}  \: \: \: \:  \text{if} \: \: [\Delta L_y  > 1200  \si{\: kpc \: km \: s^{-1}}] \: \&\\
                           \hspace{2cm}  [\Delta L_{z}  < \Delta L_{y}] \vspace{0.2cm} \\
        \texttt{y-tubes}  \: \: \: \:  \text{if} \: \: [\Delta L_z  > 1200  \si{\: kpc \: km \: s^{-1}}] \: \& \\
                            \hspace{2cm} [\Delta L_{z}  > \Delta L_{y}] \vspace{0.2cm} \\
        \texttt{resonant}  \: \: \text{if} \: \: [\Delta L_y  < 400  \si{\: kpc \: km \: s^{-1}}] \: \& \\
                           \hspace{2cm}   [\Delta L_z  < 400  \si{\: kpc \: km \: s^{-1}}] \vspace{0.2cm} \\
        \texttt{trapped}  \: \: \: \: \text{if} \: \: [\Delta L_y  < 750  \si{\: kpc \: km \: s^{-1}}] \: \& \\
        \texttt{(strong)} \hspace{0.5cm}   [\Delta L_z  < 750  \si{\: kpc \: km \: s^{-1}}] \vspace{0.2cm} \\
        \texttt{trapped}  \: \: \: \: \text{if} \: \: [\Delta L_y  < 1200  \si{\: kpc \: km \: s^{-1}}] \: \& \\
                           \hspace{2cm}   [\Delta L_z  < 1200  \si{\: kpc \: km \: s^{-1}}] \\
    \end{array}
\right.
\label{eq:orbs}
\end{equation}
Figure~\ref{fig:HS_LzLyvar} shows the result. For axisymmetric potentials (i.e. $p =1$), $\Delta L_z$ is equal to zero for all stars for, as expected, but for $p > 1$ the stars occupy different regions of the ($\Delta L_z$, $\Delta L_y$) space, meaning they are on different orbit families. While the loL stars are mainly on $z$-tube orbits, i.e. they largely conserve their $L_z$, the hiL stars are mainly resonantly trapped or are on $y$-tube orbits. The subclump stars are naturally also either resonantly trapped or on $y$-tube orbits. Hence, this figure illustrates that it is possible to have the HS stars on different orbit families in the same Galactic potential, even though they belong to the same accreted parent satellite (see also Appendix~\ref{app:freqs}).

Following our findings, we require the hiL stars to be on $y$-tube orbits, or possibly being resonantly trapped, we require the loL stars to be on the $z$-tube orbits, and we require the subclump to be strongly resonantly trapped on the $\Omega_{\phi} : \Omega_z$ = 1:1 resonance. The percentages in Fig.~\ref{fig:HS_LzLyvar} indicate the percentage of the hiL, loL and subclump stars that are on the desired orbit family. The range of $p$ and $q$ for which the percentage of hiL stars on $y$-tube orbits, the percentage of loL stars on $z$-tube orbits and the percentage of subclump stars on resonantly trapped orbits is larger than 50\% is for $(p, q) = (1.01, 1.17-1.21)$ and $(p, q) = (1.02, 1.18-1.20)$. For even larger values of $q$, an increasing number of loL stars end up on $y$-tube orbits, while for smaller $q$ than this range a decreasing number of subclump stars are resonantly trapped. For $p$ larger than this range, either the majority of subclump stars are not resonantly trapped or the majority of loL stars are on $y$-tube orbits.

\subsection{A simulated Helmi Streams-like progenitor in a mildly triaxial potential}
\label{sec:analysis:sims}

We now investigate whether a simulated HS-like progenitor on a HS-like orbit would develop the structure seen (the hiL, loL clump) over a reasonable timescale in Galactic potentials for a range of $p$ and $q$. To obtain realistic simulated HS-like phase-space positions that resemble the observations, we use \texttt{Progenitor~4} of the set of four simulated HS-like dwarf galaxies by \cite{Koppelman2019Helmi}. This progenitor has two components: stars and dark matter. The star particles follow a Hernquist profile with a stellar mass of $10^8 M_{\odot}$ and a scale radius of 0.585~kpc. The DM halo particles follow a truncated NFW profile with a total mass of $3.62 \cdot 10^9 M_{\odot}$.

To allow a good comparison to observations, we randomly select 319 star particles (the same number as stars observed in the HS) that belong to the core of  \texttt{Progenitor~4}, which we define as being the 73\% most bound star particles. We place the particles on a HS-like orbit by re-centring them to the position and velocity of a central HS star, which has $(E, L_z, L_{\bot}) = (-116246 \si{\: km^2 \: s^{-2}}, 1272 \si{\: kpc \: km \: s^{-1}}, 1871 \si{\: kpc \: km \: s^{-1}})$ and thus lies in the middle of the HS's IoM distribution. The particles have an initial $(L_z, L_{\bot}$) distribution that is positively correlated, meaning that particles with a larger $L_z$ generally have a larger $L_{\bot}$, and there is of course no gap present. We treat the particles as test particles and integrate their orbits in potentials with a range of  $1.13 < q < 1.23$ and $1.00 < p < 1.04$, motivated by the results reported in the previous sections, over a timescale of $0 - 8 $~Gyr, where the upper limit is motivated by literature estimates for the accretion time of the HS \citep{Kepley2007, Koppelman2019Helmi, Naidu2022, RuizLara2022HS}, though these estimates need to be taken with caution given our findings (see Sect.~\ref{sec:disc:accretiontime} for a more in-depth discussion).

\begin{figure*}[htb!]
\centering
    \includegraphics[width=\hsize]{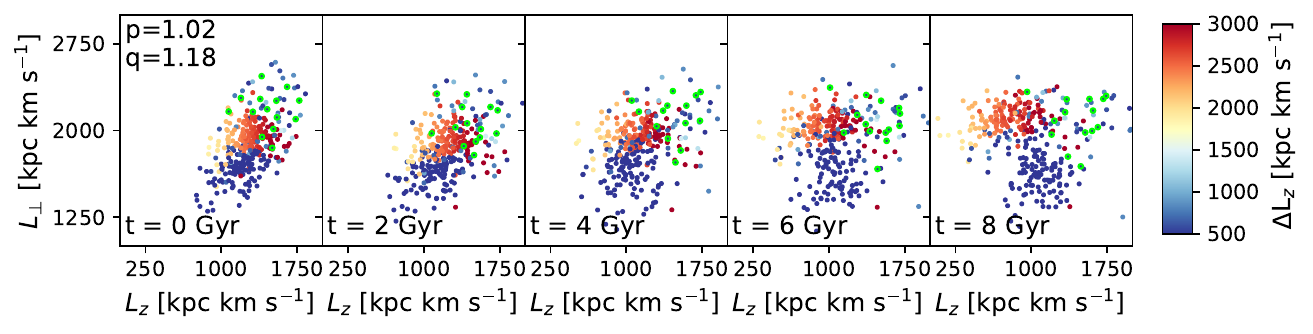}
\caption{Snapshots of the ($L_z$, $L_{\bot}$) distribution at $0, 2, 4, 6$ and $8$ Gyr of integration for 319 particles with Helmi-Stream-like phase-space positions, selected from the re-centred \texttt{Progenitor~4} by \cite{Koppelman2019Helmi}, in a potential with a triaxial NFW halo with $p = 1.02$, $q = 1.18$. The colourbar indicates the variation in $L_z$, calculated over an integration time of 100 Gyr. At $t = 0$~Gyr, the $(L_z, L_{\bot})$ distribution of the particles is positively correlated. The particles with $\Delta L_z \gtrsim 2000$ are on $y$-tube orbits, meaning their $L_z$ is no longer conserved, making them move towards lower $L_z$ in time with respect to the particles on $z$-tube orbits, which have $\Delta L_z \lesssim 1000$. Particles with high initial $L_{\bot}$ are strongly trapped by the $\Omega_{\phi}:\Omega_z$ = 1:1 resonance (selected as $\Delta L_{y,z} \lesssim 750$), and are encircled in green. A video, showing the behaviour of the particles in $(L_z, L_{\bot})$ space over time, can be found \href{https://hannekewoudenberg.github.io/helmi_streams/}{here}.
\label{fig:ICHS_time}}
\end{figure*}

We inspect the behaviour of the particles in ($L_z$,~$L_{\bot}$) space over time. Of course, the particles' orbits depend on exactly how the progenitor was re-centred, but the behaviour remains qualitatively similar for re-centrings to different central HS stars. Figure~\ref{fig:ICHS_time} shows the distribution in ($L_z, L_{\bot}$) for $q = 1.18$, $p = 1.02$ for different snapshots and illustrates how the two clumps develop and a gap is formed over time. We see how particles on $y$-tube orbits separate themselves from the particles on $z$-tube orbits as their $\Delta L_z$ is large. Particles that are resonantly trapped (in green), which have $\Delta L_{y,z} < 1200  \si{\: kpc \: km \: s^{-1}}$, are located at $(L_z, L_{\bot})$ values that are roughly consistent with those of the subclump. The final ($L_z$,~$L_{\bot}$) distribution at 8~Gyr resembles the observed HS distribution, showing two separated clumps of stars whose relative orientation is also roughly reproduced. This separation remains if we convolve the distribution with the expected observational errors.

\begin{figure*}[t!]
\centering
    \includegraphics[width=\hsize]{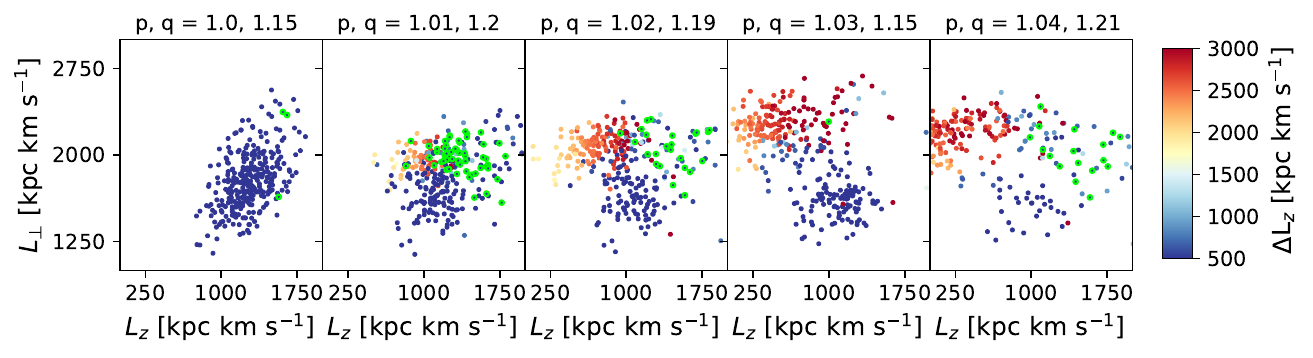}
\caption{($L_z$, $L_{\bot}$) distribution after  $8$~Gyr of integration time of 319 particles with Helmi-Stream-like phase-space positions, selected from the re-centred \texttt{Progenitor~4} by \cite{Koppelman2019Helmi}, in a potential with a triaxial NFW halo for a few combinations of $p$ and $q$. Figure~\ref{fig:ICHS_all} in Appendix~\ref{appendix:sims_pq} shows this plot for an extended grid of $p$ and $q$ values. The colourbar indicates the variation in $L_z$ over an integration time of 100~Gyr. In the axisymmetric potential, $p =1.00$, $L_z$ is an IoM and all particles are on $z$-tube orbits. For $p > 1$, the particles with larger $L_{\bot}$ are on $y$-tube orbits, meaning their $L_z$ is no longer conserved, making them move towards lower $L_z$ in time with respect to the particles on $z$-tube orbits. The timescale over which this happens depends on the degree of triaxiality of the potential, and in particular the value of $p$. A number of particles are strongly trapped by the $\Omega_{\phi}:\Omega_z$ = 1:1 resonance (selected as $\Delta L_{y,z} \lesssim 750$), and are encircled in green. A video, showing the behaviour of the particles in $(L_z, L_{\bot})$ space for a range of $p$ and $q$ over time, can be found \href{https://hannekewoudenberg.github.io/helmi_streams/}{here}.
\label{fig:ICHS}}
\end{figure*}

In Fig.~\ref{fig:ICHS} we show different distributions after 8~Gyr of integration time for a few combinations of $p$ and $q$ (and Fig.~\ref{fig:ICHS_all} in Appendix~\ref{appendix:sims_pq} shows the distributions after 8~Gyr for the entire range $1.13 < q < 1.23$ and $1.00 < p < 1.04$). For $p=1$, the particles remain as a single clump, since in an axisymmetric potential they are all on $z$-tube orbits which have constant $L_z$. Instead, for $p>1$, we find ($L_z$,~$L_{\bot}$) distributions that resemble the HS' observations, with the cases of $1.18 < q < 1.21 $ and $p \sim 1.02$ matching the most convincingly, showing two separate clumps of stars. For $p > 1.02$, two clumps can also be created, but the extent in  ($L_z$,~$L_{\bot}$) is larger than the range covered by the HS' observations, and this extent grows as $p$ gets larger. This is because $L_z$ changes faster in a more triaxial potential. This means that there is a degeneracy between the accretion time and the value of $p$, as for a more recent (earlier) accretion time, a higher (lower) value of $p$ gives similar ($L_z$,~$L_{\bot}$) distributions. 

The characteristics of the progenitor determine its extent in IoM space. If we re-do our test-particle experiment using the smaller and lighter \texttt{Progenitor~1} by \citet[][whose star particles follow a Hernquist profile with a stellar mass of $5~\cdot~10^6~M_{\odot}$ and a scale radius of 0.164~kpc, and its DM halo particles follow a truncated NFW profile with a total mass of $1.9~\cdot~10^8 M_{\odot}$]{Koppelman2019Helmi}, we similarly find that particles can be on different orbit families, but the extent in IoM space of the HS is not reproduced being too small. This confirms that \texttt{Progenitor~4} provides a better description of the HS. 

In summary, this experiment shows that a HS-like progenitor can evolve from a single clump in $(L_z, L_{\bot}$) space to two clumps over time because of the local orbital structure of the Galactic potential. Consequently, this should allow us to place constraints on the inner DM halo's shape of the Milky Way using the observed HS' dynamics.

\section{A constraint on the Milky Way's Galactic potential}
\label{sec:results}

We can strongly constrain the effective Galactic potential, and thus the potential's characteristic parameters, in the region of phase-space probed by the HS, as the observable properties of the Streams require a specific local orbital structure. That is what we set out to do in this section.

In Sect. \ref{sec:analysis:differentorbfams} we found that in our Galactic potential models with $(p, q) = (1.01, 1.17-1.21)$ and $(p, q) = (1.02, 1.18-1.20)$ the HS stars are on orbits such that the formation of the HS clumps may be explained. However, this range of $(p,q)$ values was obtained for a potential whose parameters were all fixed except for the shape of the DM halo. To provide a more robust estimate including uncertainties and to explore the influence of degeneracies, we now proceed to vary the disc mass, halo scale radius, halo density and $p$ and $q$ while fitting the rotation curve data by \cite{Eilers2019} and \cite{Zhou2023}, following the method outlined in Appendix~\ref{app:pot}. We simultaneously maximise the number of subclump stars that are strongly resonantly trapped by the $\Omega_{\phi}:\Omega_z$ = 1:1 resonance, which we translate into the requirement $\Delta L_{y, sub}  < 750 \si{\: kpc \: km \: s^{-1}} $ and $\Delta L_{z, sub}  < 750 \si{\: kpc \: km \: s^{-1}} $. We focus on this resonance as it is pivotal in setting the local orbital structure, and it is required to explain the existence of the kinematically cold subclump. We run a Monte Carlo Markov Chain (MCMC) using \texttt{emcee} \citep{emcee} with 40 walkers and 4000 steps. By performing the MCMC, we can estimate how much the location of the resonance shifts by varying parameters of the potential. Moreover, the MCMC can sample the parameter space more finely than we have done so far. Our likelihood has the form
\begin{equation}
    \mathcal{L}(\vec{\theta}) = - 101 + 100 \cdot \dfrac{N_{\rm{sub, \: strongly \: trapped}}}{N_{\rm{sub, \: tot}}}  - \dfrac{1}{2 N_{\rm{RC}}} \sum_{i=1}^{N_{\rm{RC}}} \left(  \dfrac{v_{\rm{c}, \it{i}}^{\rm{d}} - v_{\rm{c}}^{\rm{m}}}{\sigma_{v_{\rm{c}, \it{i}}}} \right)^2
    \label{eq:likelihood}
\end{equation}
\noindent where the superscript d indicates the data and m the model. The two likelihood terms have a relative importance of about $10:1$, respectively. We set a simple flat prior and allow
\begin{equation}
P(\vec{\theta}) = \left\{
    \begin{array}{ll}
        1 & \text{if} \left\{
            \begin{array}{ll}
                1.0 \leq p \leq 1.1  \\
                0.7 \leq q \leq 1.4  \\
                10 \leq r_{\rm{h}} \leq 30 \: \text{kpc} \\
                10^{5} \leq \rho_{\rm{0, h}} \leq 10^{8} \: M_{\odot} \text{~kpc}^{-3} \\
                3.5 \cdot 10^{10} \leq M_{\rm{discs}} \leq 5.5 \cdot 10^{10} M_{\odot}
            \end{array}
            \right.  \\
        0 & \text{otherwise.}
    \end{array}
\right.
\label{eq:priorMCMC}
\end{equation}
\noindent and take $\vec{\theta}_{i} = [p_i, q_i, r_{\rm{h},\it{i}}/ {\rm kpc}, \, \rho_{\rm{0,h},\it{i}}/(M_{\odot}~\text{~kpc}^{-3}), \, M_{\rm{discs},\it{i}}/M_{\odot}] = [1.02, 1.20, 15.7, 10 \cdot 10^{6}, 4.5 \cdot 10^{10}]$ as an initial guess, motivated by our findings and the parameters of the fiducial potential. Recall that we have fixed the discs scale-lengths and relative density near the Sun (see Appendix~\ref{app:pot}), which is why we consider the sum of the masses of the thin and thick discs, $M_{\rm{discs}}$, as the free parameter.

\begin{figure}[t!]
\centering
    \includegraphics[width=0.5\textwidth]{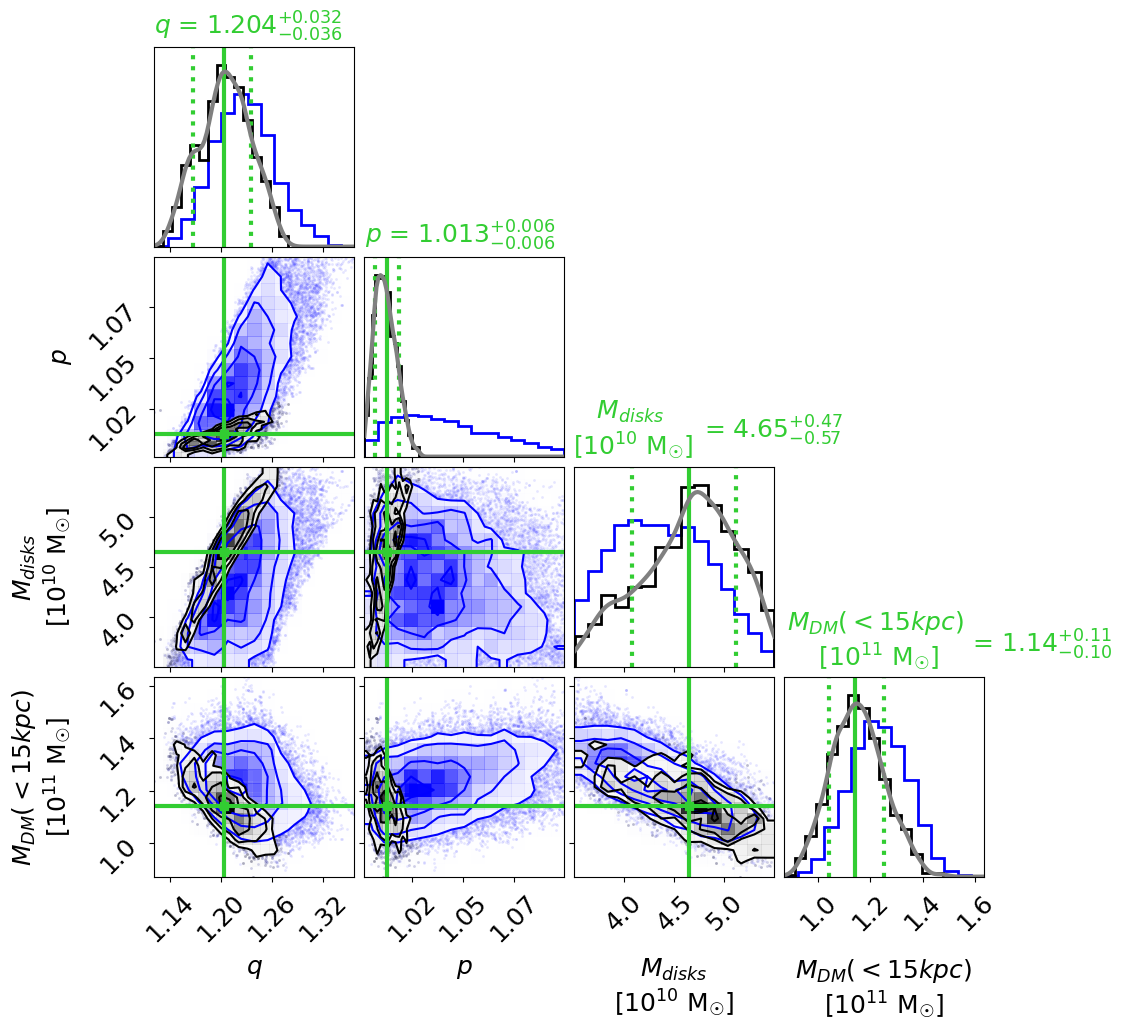}
\caption{Posterior distribution of the parameters $q$, $p$, $M_{\rm{discs}}$ and $M_{\rm{DM}}$(<15kpc) that maximises the number of subclump stars on strongly resonantly trapped orbits and fits the rotation curve data by \cite{Eilers2019} and \cite{Zhou2023}, in blue. MCMC steps of walkers that got stuck have been removed. The posterior parameter distribution shown in black corresponds to potentials in which at least 40\% of the loL stars are on $z$-tube orbits. The peak value of this distribution, which we take as our best estimate, is indicated in green solid lines in all panels, and was determined using a kernel-density estimation. The 31.73\% (68.27\%)  percentile of the data lower (higher) than the peak value is taken as the lower (upper) 1$\sigma$ uncertainty and are indicated with dotted green lines in the panels showing histograms of the marginalised distributions. The grey lines in those same panels show the kernel-density estimation of the marginalised distributions.
\label{fig:cornerpq_zoom}}
\end{figure}

\subsection{Results}
\label{sec:results:results}

The MCMC converged, though with a relatively low acceptance fraction of $\sim$0.27. We have removed the first 1000 steps as burn-in and steps of walkers that got stuck. Figure~\ref{fig:cornerpq_zoom} shows the posterior parameter distribution for $p$, $q$, $M_{\rm{discs}}$ and $M_{\rm{DM}}$(<15kpc) in blue, where $M_{\rm{DM}}$(<15kpc) is computed by integrating Eq. \ref{eq:NFW} for a given $r_{\rm{h}}$, $\rho_{\rm{0,h}}$, $p$ and $q$ within an ellipsoid which has a minor axis of 15 kpc and axis ratios set by $p$ and $q$. We constrain ourselves to $15~$kpc as this is the limit of our rotation curve data, and thus show $M_{\rm{DM}}$(<15kpc). In this context, the degeneracy between $M_{\rm{discs}}$ and $M_{\rm{DM}}$(<15kpc) seen in Fig.~\ref{fig:cornerpq_zoom} is not surprising. The other degeneracies between the parameters can easily be explained as well. The degeneracy between the disc mass and $q$ reflects the fact that the required local orbital structure constrains the total effective potential flattening. For a larger disc mass, a larger $q$ is required to compensate for the flattening introduced by the disc. Following a similar reasoning, a more massive DM halo implies a lower~$q$, though this degeneracy is weaker. The degeneracy between $p$ and $q$ reflects the fact that the $\Omega_{\phi} : \Omega_z$ = 1:1 resonance is located on a diagonal line in $p$ and $q$, see Fig.~\ref{fig:mapstogether} and Fig.~\ref{fig:HS_LzLyvar}. In almost all unique potentials of the chain, $\sim 99.6 \%$, all nine subclump stars are strongly resonantly trapped. In more than 22\% of the potentials, 5 or more subclump stars are on the resonant orbit while the rest is strongly resonantly trapped.

Though this posterior parameter distribution beautifully reflects where the subclump is resonantly trapped, it does not contain information about the orbits of the hiL and the loL stars. We find that in all potentials, the hiL stars are either resonantly trapped or on the $y$-tube orbits, as desired, but in a majority of the potentials, especially for larger $p$, most loL stars are on $y$-tube orbits as well. Therefore, we analyse the orbits of the loL stars in each sampled potential, and add the constraint that at least 40\% of the loL stars are on $z$-tube orbits. This constrained posterior parameter distribution is shown in black in  Fig.~\ref{fig:cornerpq_zoom} and shows that a low $p$ is required, in concordance with Fig.~\ref{fig:HS_LzLyvar} (though in that case only $p$ and $q$ were varied).

We proceed now to determine our best-fit model. An inspection of Fig.~\ref{fig:cornerpq_zoom} shows that the posterior distribution of $p$ is skewed and that parameters are correlated. Hence, the median or average values of the marginalised distributions do not reflect the peak of the posterior parameter distribution. Therefore, to obtain the Galactic potential model that satisfies the likelihood best, we select the global peak value of the posterior distribution. Next, we use \texttt{scipy.stats.gaussian\_kde} to obtain a kernel-density estimate of the posterior distribution, where we use the default settings but with a bandwidth (which scales the width of the kernel) equal to $2/3$ \texttt{'scott'} to capture the finer details of the distribution\footnote{The obtained peak values are consistent when the default bandwidth \texttt{'scott'} or \texttt{'silverman'} is used, which are determined using Scott's Rule and Silverman’s Rule, respectively. For a bandwidth equal to $1/2$ \texttt{'scott'}, a local maximum instead of the global maximum is obtained.}. The peak values are those corresponding to the highest density and are shown with solid green lines in Fig.\ref{fig:cornerpq_zoom}. Given the peak values, we then take the 31.73\% (68.27\%)  percentile of the data lower (higher) than the peak value as the lower (upper) 1$\sigma$ uncertainty. These are indicated with dashed green lines in Fig.\ref{fig:cornerpq_zoom}. Hence, we find $p=1.013^{+0.006}_{-0.006}$, $q=1.204^{+0.032}_{-0.036}$, $M_{\rm{discs}}=4.65^{+0.47}_{-0.57}\cdot10^{10} M_{\odot}$ and $M_{\rm{DM}}(< 15 \text{kpc})=1.14^{+0.11}_{-0.10}\cdot10^{11} M_{\odot}$. In the best-fit potential, 5 subclump stars are on the resonance, while 4 are strongly resonantly trapped. All hiL stars are either on the resonance (9\%), resonantly trapped (57\%) or on $y$-tube orbits (21\%), and 80\% of the loL stars are on $z$-tube orbits. This is in agreement with our findings of Fig.~\ref{fig:HS_LzLyvar}.
\section{Discussion} 
\label{sec:disc}

\subsection{Comparison to the literature}

In this work, we have constrained the Milky Way's DM halo shape and characteristic parameters, and the mass of its discs. This is the first constraint on the degree of triaxiality of the DM inner halo and it is  based on the dynamics of phase-mixed streams. Here, we compare the Galactic potential characteristics that we found to the literature. 

We constrained the discs' total mass $M_{\rm{discs}}=4.65^{+0.47}_{-0.57}\cdot10^{10} M_{\odot}$. Given that we assumed a fixed local density ratio and fixed parameters for the scale length and scale height, this implies $M_{\rm{thin}} = 3.12^{+0.31}_{-0.38} \cdot 10^{10} M_{\odot}$ and $M_{\rm thick} = 1.53^{+0.15}_{-0.19} \cdot 10^{10} M_{\odot}$. This is consistent with \cite{Bland-Hawthorn2016} and references therein. We also constrained the total DM halo mass within 15~kpc to be $M_{\rm{DM}}(< 15 \text{kpc})=1.14^{+0.11}_{-0.10}\cdot10^{11} M_{\odot}$. This corresponds to a local DM density of $\rho_{0, h}(R_{\odot}, z_{\odot}) = 8.3^{+1.1}_{-0.8}\cdot 10^{-3} ~M_{\odot}$~pc$^{-3}$ or $0.32^{+0.04}_{-0.03}~\text{GeV}$~cm$^{-3}$, which is consistent with the local DM densities of the models of \cite{McMillan2017} and \cite{Cautun2020}, but also with the compilation of estimates presented in \cite{Read2014} and \cite{Bland-Hawthorn2016}. The total mass within 15~kpc is $M_{\rm{tot}}(<15 \: \text{kpc}) = 1.80^{+0.08}_{-0.07} \cdot 10^{11} M_{\odot}$.

The posterior parameter distribution for the DM halo scale radius $r_{\rm{h}}$ reveals that a value below the lower bound of 10~kpc would be preferred. This could indicate that a different DM profile is more suitable, possibly a contracted NFW \citep{Cautun2020}. We note that \cite{Ou2024} have suggested in their fit to recent rotation curve data that a cored Einasto profile provides a better fit over a (generalised) NFW profile. Despite being cored, their Einasto profile is relatively steep with an $r_{-2} = 9.17$~kpc, which is very comparable to the value we obtain (as $r_{\rm{h}} = r_{-2}$ for an NFW profile).

Similar to the fiducial potential presented earlier (and discussed in Appendix~\ref{app:pot}), the circular velocity at the position of the Sun is $233.3^{+2.8}_{-2.1} \: \si{km \: s^{-1}}$. Also the model's rotation curve, shown in Fig.~\ref{fig:RC_fit}, agrees with the rotation curve of the fiducial potential. Lastly, we confirm that our model agrees with the determination of the vertical force 1.1~kpc from the Galactic plane as a function of radius by \cite{Bovy2013}.

\begin{figure}[t!]
\centering
    \includegraphics[width=\hsize]{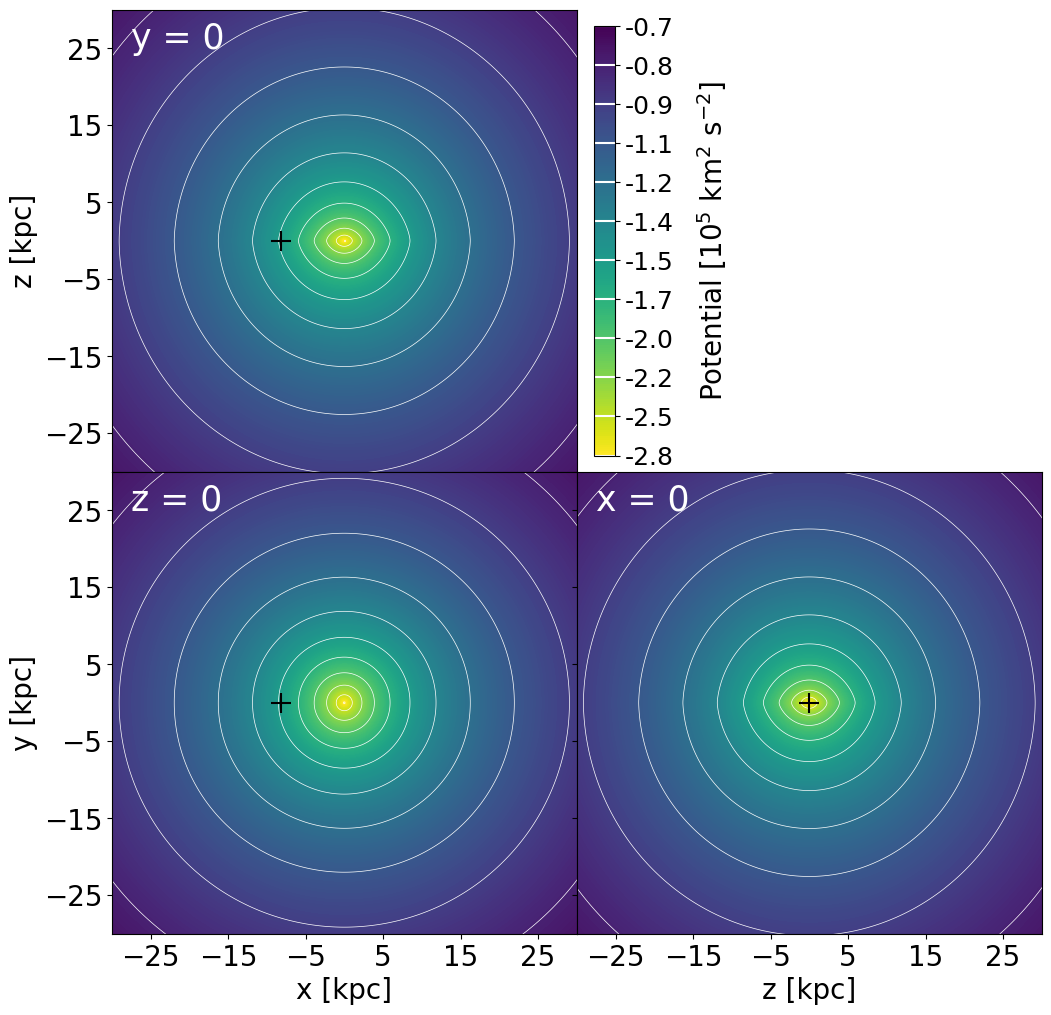}
\caption{Isopotential contours of the effective potential for the best-fit model of this work, on the plane $y = 0$ (top left), $z = 0$ (bottom left), and $x = 0$ (bottom right). For reference, the position of the Sun is indicated with a black plus sign. The effective potential transitions to roughly spherical at $r \sim 15$~kpc. 
\label{fig:isopotential}}
\end{figure}

\begin{figure*}[htb!]
\centering
    \includegraphics[width=\hsize]{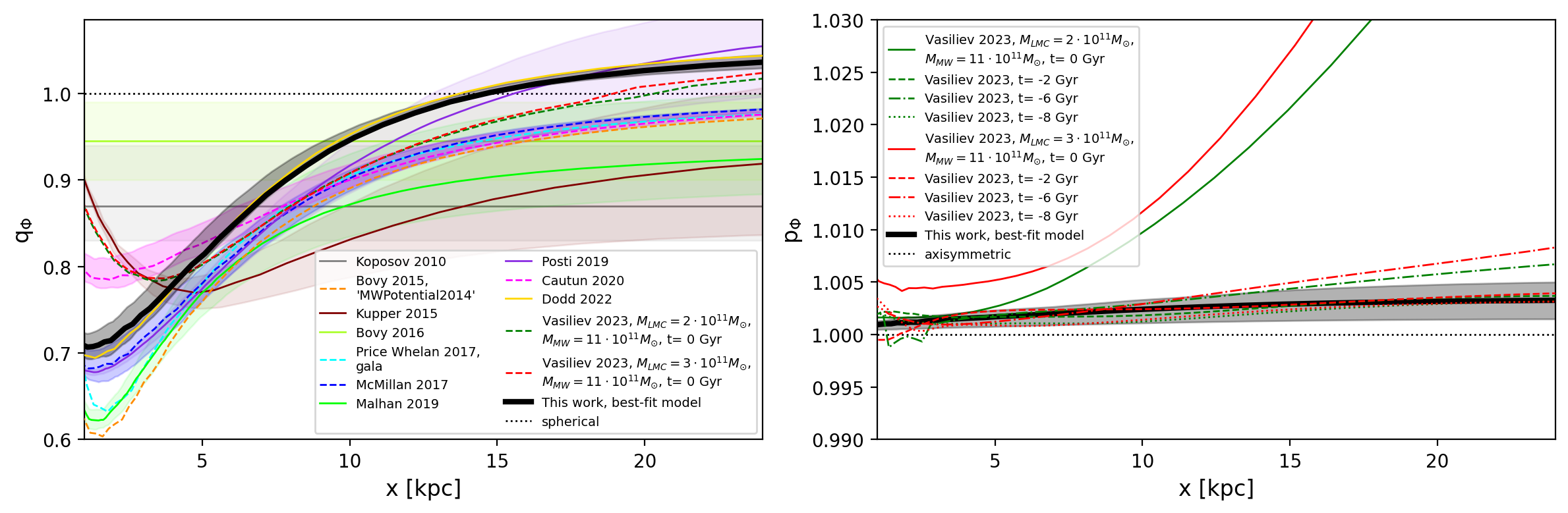}
\caption{Overview of the flattening of the effective potential $q_{\Phi}$ and $p_{\Phi}$, as a function of Galactocentric distance along the $x$-axis for different analytical Milky Way mass models, namely those by \cite{Bovy2015galpy}, \cite{softwarecitegala}, \cite{McMillan2017}, \cite{Cautun2020},  and Milky Way mass models constrained by stellar streams, namely \cite{Koposov2010} \cite{Kupper2015}, \cite{Bovy2016GD1Pal5} and \cite{Malhan2019}, by globular cluster dynamics \citep{Posti2019} and by previous work on the HS \citep{Dodd2022}. This figure also shows two Galactic potential models by \cite{Vasiliev2023_LMC} with the $\mathcal{L}3, \mathcal{M}11$ and $\mathcal{L}2, \mathcal{M}11$ halo models at the present day in the left panel, and at different moments in time in the right panel. The black line shows the results obtained in this paper. The uncertainties derived for \cite{McMillan2017}, \cite{Posti2019}, \cite{Cautun2020} and this work have been obtained by randomly sampling 100 potentials from the respective chains, while for \cite{Kupper2015} and \cite{Malhan2019} we have sampled within the uncertainties quoted in those works. In all cases, we have taken the 16$^{\rm th}$ and 84$^{\rm th}$ percentile values. For \cite{Koposov2010} and \cite{Bovy2016GD1Pal5} we show the uncertainties quoted in those works. Models that are based on constraints of the DM halo shape are shown with solid lines, models that assume a spherical DM halo are shown with dashed lines, and their similarity is apparent in $q_{\phi}$ at larger distance, except for the \cite{Vasiliev2023_LMC} models, which take into account the infall of the LMC. The differences at small distances are due to varying assumptions on the discs and bulge, but are not relevant for the work presented here.
\label{fig:flattening_funcR}}
\end{figure*}

\begin{figure}[htb!]
\centering
    \includegraphics[width=\hsize]{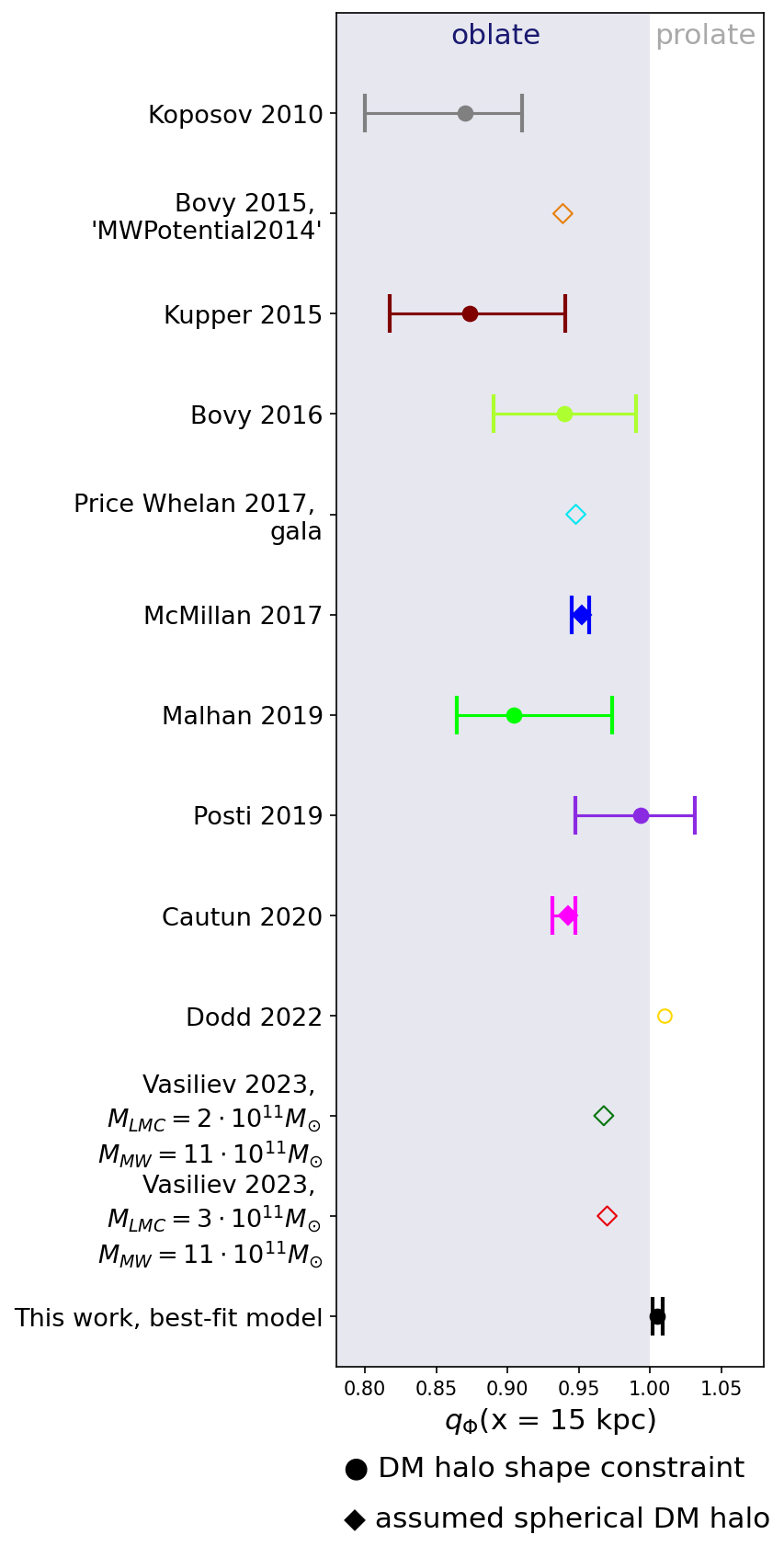}
\caption{Overview of the flattening of the effective potential $q_{\Phi}$ at Galactocentric coordinate $x = 15$~ kpc of different analytical Milky Way mass models, namely those by \cite{Bovy2015galpy}, \cite{softwarecitegala}, \cite{McMillan2017}, \cite{Cautun2020},  and Milky Way mass models constrained by stellar streams, namely \cite{Koposov2010}, \cite{Kupper2015}, \cite{Bovy2016GD1Pal5} and \cite{Malhan2019}, globular cluster dynamics \citep{Posti2019} and the HS \citep{Dodd2022}. This figure also shows two Galactic potential models by \cite{Vasiliev2023_LMC} with the $\mathcal{L}3, \mathcal{M}11$ and $\mathcal{L}2, \mathcal{M}11$ halo models. Models that are based on constraints of the DM halo shape are shown with circles, models that assume a spherical DM halo are shown with diamonds. Solid symbols indicate models for which uncertainties on the effective potential could be derived, as described in the caption of Fig. \ref{fig:flattening_funcR}. Open symbols indicate that no uncertainty on the potential parameters was available, and hence no uncertainty on the effective potential flattening could be derived. 
\label{fig:flattening_R15} }
\end{figure}

The HS constrain the shape and mass distribution of the effective Galactic potential, which is obtained by adding the contributions of all Galactic components. Fig.~\ref{fig:isopotential} shows the isopotential contours of the effective potential of our best-fit model in different planes. To place our findings in context, we compare our results to the effective flattening of Galactic potential models from the literature. We evaluate the shape of all these models by approximating the effective isopotential at different radii by a triaxial ellipsoid and determining the axis ratios. Fig.~\ref{fig:flattening_funcR} shows $q_{\Phi}$ and $p_{\Phi}$ as a function of Galactocentric coordinate $x$, while Fig.~\ref{fig:flattening_R15} shows $q_{\Phi}$ at a fixed value, $x = 15$~kpc. The comparison in the innermost regions, $x < 5$~kpc, is not very meaningful as the curves differ due to different assumptions on the discs, bulge and halo profile (e.g. \cite{Cautun2020} assumes a contracted DM halo, and \cite{Vasiliev2023_LMC} assumes a heavier bulge). The constraints obtained from fitting spatially coherent streams result most often in oblate $q_{\phi}$-values that agree within uncertainty with each other. For $x > 10$~kpc most models assume a spherical DM halo, and as a consequence 
the resulting effective potentials are oblate in the region occupied by the HS. Instead, the effective flattening of our model turns prolate for $x \gtrsim 13$~kpc. Our findings agree with the work by \cite{Dodd2022} and \cite{Posti2019}, and also \cite{Vasiliev2023_LMC}'s models turn prolate at $x \sim 18$~kpc.

As the HS constrain the effective potential, we cannot distinguish with certainty whether the measured triaxiality is due to the DM halo's shape or to the effect of a perturbation such as the LMC. We can however neglect the triaxiality induced by the Galactic bar and spiral arms, as these do not strongly affect the dynamics of the HS since they orbit high above the plane (see Fig.~\ref{fig:orbparams}). To investigate the effect of the LMC in the region probed by the HS, we compare $p_{\Phi}$ of our model to the $\mathcal{L}3, \mathcal{M}11$ and $\mathcal{L}2, \mathcal{M}11$ time-dependent Galactic potential models by \cite{Vasiliev2023_LMC}. In these two models, the LMC is on its second passage around the Milky Way. Both models have a MW halo with a virial mass of $M_{\rm{vir, MW}} = 11 \cdot 10^{11} M_{\odot}$, and have $M_{\rm{LMC}} = 3 \cdot 10^{11} M_{\odot}$ and $M_{\rm{LMC}} = 2 \cdot 10^{11} M_{\odot}$, respectively \citep{Vasiliev2023_LMC}. The right panel of Fig.~\ref{fig:flattening_funcR} shows that the present-day perturbation of the LMC exceeds the triaxiality required by the HS, with $1 < p_{\Phi} < 1.06$. However, at earlier times, the perturbation is of the same order as the required triaxiality. Although beyond the scope of this work, it could be worthwhile to investigate the behaviour of the HS orbits in a potential including both the LMC and a prolate DM halo. The \cite{Vasiliev2023_LMC} models assume a spherical NFW DM halo as a starting point, and as a consequence, all HS stars are placed on $z$-tube orbits and in $(L_z, L_{\bot})$ space the Helmi Streams' clumps do not remain separated in time. 

\subsection{Limitations, uncertainties and systematics}

A number of simplifying assumptions have been made throughout this work. To start, we assumed that the DM halo follows an NFW profile with a triaxial shape whose axes are aligned with those defined by the MW disc. Recent work by \cite{Han2023,Han2023sims} has however suggested that the Milky Way's DM halo might be titled. Besides this, there is evidence that the Milky Way's DM halo could be contracted \citep[see e.g.][]{Dutton2016, Cautun2020} and follows a mass distribution whose shape may vary with radius \citep{VasilievTango2021, Shao2021, Cataldi2023}. Next, in all analyses performed in this work we have assumed a static potential. However, the effect of the LMC's perturbation could prove important, as discussed in the previous section\footnote{The MW is also being perturbed by the ongoing merger with the Sagittarius dwarf galaxy \citep{Ibata1994}. However, we expect that it is unlikely that this led to the formation of the two clumps, as all stars will have felt the perturbation equally independent of their orbital phase. Interestingly, though, the HS and Sagittarius have similar orbital frequency ratios, possibly suggesting a connection, such as group infall \citep{Li2009}.}. It would also be interesting to explore the local orbital structure for non-parametrised forms of the Galactic potential,  using for example (low-order) basis function expansions \citep[see e.g.][]{GaravitoCamargo2021, Lilleengen2022, Vasiliev2023_LMC}. 

The peculiar motion of the Sun also introduces a systematic uncertainty on the estimated values of $p$ and $q$. A larger (smaller) value or the velocity of the Sun could lead to an overcorrection (undercorrection) when transforming from observables to Galactocentric coordinates. This means that we add (subtract) velocity to (from) the true motion of the stars, which in turn will make them reach farther (less far) out. In general, the farther away one is from the Galactic Centre, the more the effective potential becomes dominated by the DM halo (see also Fig.~\ref{fig:relative_Mr} in Appendix~\ref{app:pot}). Hence, if all HS stars would have larger apocentres, the DM halo's value of $q$ that is needed to reach an effective flattening of $q_{\Phi} = 1$ would be smaller. We use the subclump stars to obtain an order of magnitude estimate of this effect. For these stars, a difference in $v_{\rm{LSR}}$ and thus $V_{\odot}$ of $\sim 7 \si{\: km \: s^{-1}}$ leads to a shift in apocentre of $\sim 2$~kpc, while the pericentres remain similar. This in turns shifts the location of the $\Omega_{\phi} :\Omega_z$ = 1:1 resonance in $q$ by about $\sim0.03$. Hence, the systematic uncertainty associated to this effect is $\sigma_{\rm{sys}} \sim 0.03$, which is very small. 
 
\subsection{Accretion time estimate}
\label{sec:disc:accretiontime}

In the literature, the asymmetry between the numbers of stars in the HS with positive and negative $v_z$ has been used to estimate the time of accretion time  \citep{Kepley2007, Koppelman2019Helmi}.  However, such estimates will be biased towards more recent accretion times as the hiL clump, which we have established is on or close to a resonance, will phase-mix more slowly. Therefore, to estimate the time of accretion, one should use the HS' loL clump only.  The loL clump's ratio, $\dfrac{N_{\rm loL}(v_{z}^{+})}{N_{\rm loL}(v_{z}^{-})}\sim 0.62$, is significantly different from the ratio we obtain for all HS stars,  $\dfrac{N_{\rm HS}(v_{z}^{+})}{N_{\rm HS}(v_{z}^{-})} \sim 0.24$. 

\begin{figure*}[htb!]
\centering
    \includegraphics[width=0.885\hsize]{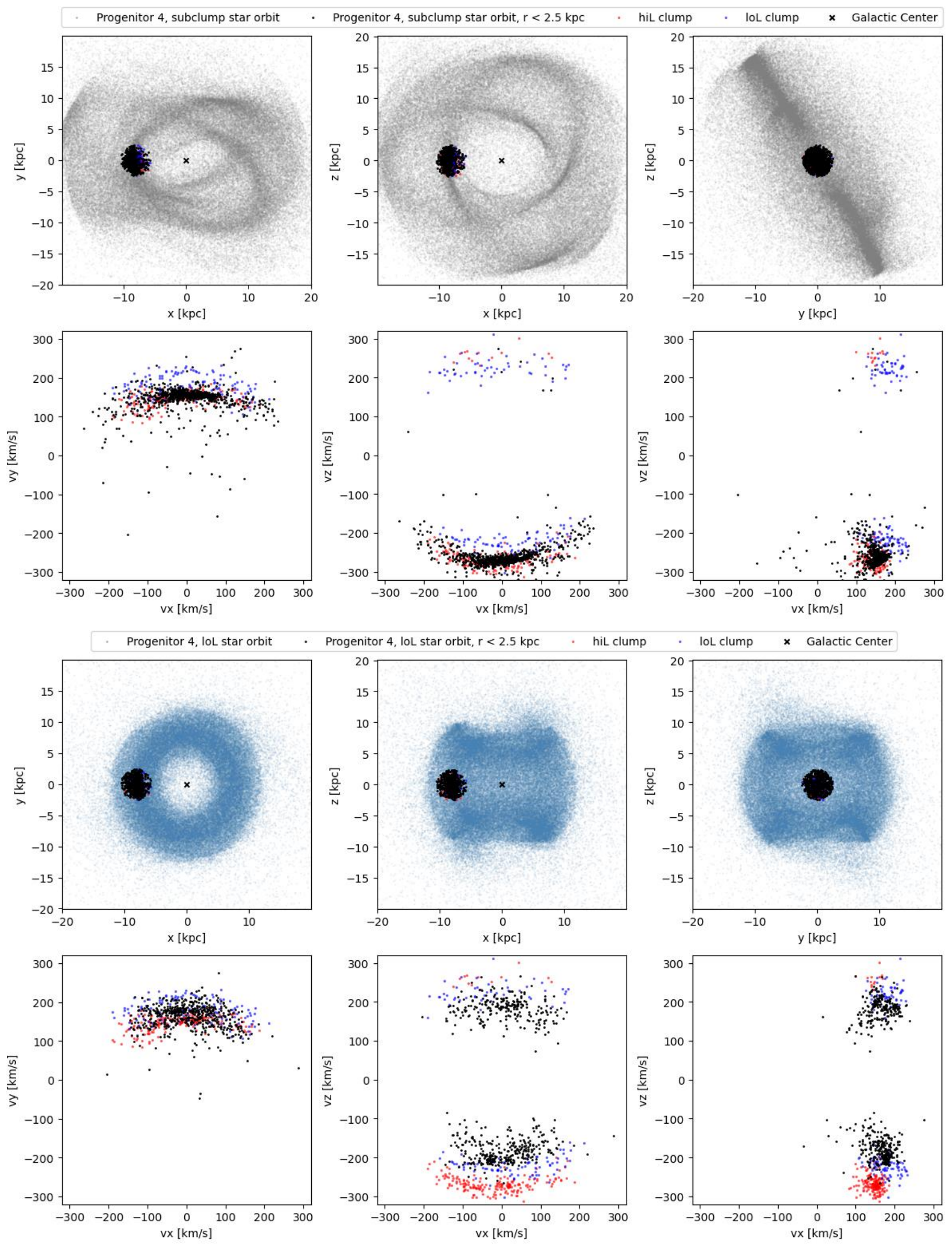}
\caption{Phase space distribution of \texttt{Progenitor~4} by \cite{Koppelman2019Helmi} on two different orbits after 8~Gyr of integration time in a potential with $p = 1.02, q = 1.19$.  Particles within a local volume selection, i.e. $d < 2.5~$kpc, are shown in black. For comparison, the hiL and loL HS stars are shown in red and blue, respectively. The top two rows show the particles of \texttt{Progenitor~4} at the present day after it has been re-centred and integrated forward from the 6D phase-space coordinate of a subclump star 8~Gyr ago, which is resonantly trapped on the $\Omega_{\phi}:\Omega_z$ = 1:1 resonance. Consequently, a large part of the \texttt{Progenitor~4} particles is resonantly trapped, as is evident from the substructure in configuration space, the flattened particle distribution in the $(y,z)$ plane and the low $\dfrac{N(v_{z}^+)}{N(v_{z}^{-})} \sim 0.01$ ratio. The bottom two rows show particles of \texttt{Progenitor~4} at the present day after it has been re-centred and integrated forward from the 6D phase-space coordinate of a loL star 8~Gyr ago, which is on a $z$-tube orbit. The particles appear more mixed, as they show less substructure in configuration space, have a more diffuse velocity distribution and a higher $\dfrac{N(v_{z}^+)}{N(v_{z}^{-})} \sim 0.7$ ratio.
\label{fig:test_particle_resnonres}}
\end{figure*}

To illustrate how a resonance affects phase-mixing time-scales, we set up a test-particle simulation experiment. We use \texttt{Progenitor~4} by \cite{Koppelman2019Helmi} in a potential with $p = 1.02$, $q=1.19$ and explore two cases. Firstly, we integrate the orbit of a subclump star, which is on the $\Omega_{\phi}:\Omega_z$ = 1:1 resonance, to an apocentre 8~Gyr ago and re-centre \texttt{Progenitor~4} to that 6D phase-space coordinate. We then integrate the orbits of all particles forward to the present day. We select particles that are in a local volume, i.e. $d < 2.5$~kpc, and inspect their positions and velocities. 
The result is shown in the top two rows of Fig.~\ref{fig:test_particle_resnonres}. A large part of the \texttt{Progenitor~4} particles is resonantly trapped, as is evident from the flattened particle distribution in the $(y,z)$ plane. We clearly see that \texttt{Progenitor~4} is not yet fully phase-mixed and its streams can still be identified as coherent structures in configuration space. Moreover, the local volume particles have a tight distribution in velocity space and a low $\dfrac{N(v_{z}^{+})}{N(v_{z}^{-})} \sim 0.01$ ratio. 
Next, we follow the same procedure but use a loL star, which is on a $z$-tube orbit, for re-centring. The result is shown in the bottom two rows of Fig.~\ref{fig:test_particle_resnonres}. The particles appear more mixed, as they show less substructure in configuration space, have a more diffuse velocity distribution and a significantly higher $\dfrac{N(v_{z}^{+})}{N(v_{z}^{-})}\sim 0.7$ ratio. We find that the velocity distribution and the $v_z$ ratio of the particles match the loL observations roughly after at least 5~Gyr of evolution, implying that, according to this experiment, the HS were likely accreted at least 5 Gyr ago. This accretion time estimate is robust to recentrings to different loL-like orbits and to forward or backward integration, as is expected for a structure on a regular, non-resonant orbit. This experiment emphasises that the degree of asymmetry in the number of stars associated to streams that have undergone some amount of phase-mixing cannot be straightforwardly used to estimate accretion times when a substructure is located close to or on a resonance (a similar conclusion seems to apply to the streams from {\it Gaia}-Enceladus, see \citealt{Dillamore2024}, although in this case, due to resonances with the Galactic bar).

An independent measure of the accretion time of the HS can be derived from the rate of variation in $L_z$ of the $y$-tube orbits. Fig.~\ref{fig:ICHS} shows that the timescale over which $L_z$ varies is linked to the value of $p$: a larger value of $p$ implies a faster change in $L_z$. Fig.~\ref{fig:ICHS} shows the $(L_z, L_{\bot})$ distribution of the particles after 8~Gyr of integration time, and $p > 1.02$ results in distributions that span a too large range in $L_z$ compared to the observations. However, the same figure but for an integration time of 4 or 6~Gyr shows two clumps in the $(L_z, L_{\bot})$ distribution with an extent resembling the HS observations for $p = 1.03$ and $p = 1.04$, respectively. What exact value of $p$ is preferred in these test-particle simulations thus depends on the accretion time of the HS. Since our MCMC results, based on the observed HS stars, indicate that $p \lesssim 1.02$, this allows us to tentatively constrain the Helmi Streams' accretion time to about at least 6~Gyr ago, confirming our earlier estimate based on the degree of phase-mixing seen for the loL streams.

\section{Conclusion}
\label{sec:conclusion}

We have studied the dynamical properties of the (phase-mixed) Helmi Streams using \textit{Gaia} DR3 data and have used it to constrain the Galactic potential within 20~kpc, the region probed by the streams stars' orbits. The dynamics of the Helmi Streams are peculiar, as their stars separate into two clumps in angular momentum space \citep{Dodd2022}, of which one has a larger total angular momentum (hiL) than the other (loL). The phase-space distribution of the Helmi Streams indicate that the hiL clump is less phase-mixed than the loL clump, as is apparent from their $v_z$ distribution. We have also confirmed that the hiL clump contains substructure in the form of a kinematically cold subclump (known as S2, see \citealt{Myeong2018}).

To explain these characteristics, we have studied the orbital structure in the volume of phase-space probed by the Helmi Streams in a range of Galactic potentials models with a triaxial NFW halo, parametrised by the (density) axis ratios $p$ and $q$. We have used the Lyapunov exponent as chaos indicator, the variation in $L_z$ and $L_y$ to classify orbits, and orbital frequencies $(\Omega_R, \Omega_{\phi}, \Omega_z)$ to identify resonances. We 
have found that for triaxial DM halos with flattening $q \sim 1.2$, the $\Omega_{\phi}:\Omega_z$ = 1:1 resonance acts as a separatrix between $z$-tube and $y$-tube orbits, which are orbits that rotate around the $z$- and $y$-axis, respectively. In the case of $p<1$, this resonance causes chaoticity, while for $p>1$, this resonance stabilises orbits and resonantly traps them. This means $p>1$ is favoured, as it is able to sustain the kinematically cold subclump. The location of the resonance in relation to the hiL and loL clumps, and therefore also the transition in orbit family, varies with $(p, q)$.  Consequently, there is a region of $(p,q)$ parameter space where the hiL clump is on a different orbit family than the loL clump, providing an explaination of the longevity and separation of the clumps.    Specifically, we have found that the hiL clump can be partially resonantly trapped, while the kinematically cold subclump is strongly resonantly trapped and the loL clump is located off-resonance, and this in turn explains their different degrees of phase-mixing. In such Galactic potential models, the hiL and loL clump with the correct dynamical characteristics can be formed over time from a single progenitor, which we have demonstrated by integrating and inspecting the orbits of simulated star particles of a Helmi Streams-like progenitor.

We have taken these findings a step further and used them to constrain the characteristics of the Galactic potential. We have performed a Monte Carlo Markov Chain sampling of Galactic potential models in which we vary $p$, $q$, the total disc (thin and thick) mass, DM halo scale radius and density, and we require the kinematically cold subclump to be resonantly trapped on the $\Omega_{\phi}:\Omega_z$ = 1:1 resonance. We also require the model to fit recent rotation curve data \citep{Eilers2019, Zhou2023}. Additionally, we require that at least 40\% of the loL stars is on $z$-tube orbits. We find  $p=1.013^{+0.006}_{-0.006}$, $q=1.204^{+0.032}_{-0.036}$, $M_{\rm{discs}}=4.65^{+0.47}_{-0.57}\cdot10^{10} M_{\odot}$ and $M_{\rm{DM}}(< 15 \text{kpc})=1.14^{+0.11}_{-0.10}\cdot10^{11} M_{\odot}$. Our best-fit model has $v_{\rm{c}}(R_{\odot})=233.3^{+2.8}_{-2.1}\si{km \: s^{-1}}$. In the best-fit potential, 5 subclump stars are on the resonance, while 4 are strongly resonantly trapped. All hiL stars are either on the resonance (9\%), resonantly trapped (57\%) or on y-tube orbits (21\%), and 80\% of the loL stars are on z-tube orbits. Our constraint on the shape of the DM halo is extremely strong. This can be can be explained by the fact that we combine precise rotation curve data, which strongly constrains the mass enclosed on the midplane of the Galaxy, with a constraint on the local orbital structure of the potential, which depends on the effective shape and characteristics of the potential. 

Our findings suggest that the orbital structure of the potential can strongly influence the dynamical properties of substructures in the Galaxy, especially if these are located near a separatrix. Among the possible consequences, it may complicate the identification of debris due to chaos which accelerates phase-mixing \citep{PriceWhelan2016, Yavetz2022}. Furthermore the determination of the accretion time may also affected due to the presence of resonances which in contrast slows down the process of phase-mixing. On the other hand, it gives us a powerful and exciting new tool to probe the orbital structure of the potential, and hence its characteristic parameters \citep{Caranicolas2010, Zotos2014}, 
as is shown in this work. 


\begin{acknowledgements}
This research has been supported by a Spinoza Grant from the Dutch Research Council (NWO). HCW thanks Lorenzo Posti and Paul McMillan for providing the chains of their MCMC runs.
This work has made use of data from the European Space Agency (ESA) mission
{\it Gaia} (\url{https://www.cosmos.esa.int/gaia}), processed by the {\it Gaia}
Data Processing and Analysis Consortium (DPAC,
\url{https://www.cosmos.esa.int/web/gaia/dpac/consortium}). Funding for the DPAC
has been provided by national institutions, in particular the institutions
participating in the {\it Gaia} Multilateral Agreement.

Throughout this work, we've made use of the following packages: \texttt{astropy} \citep{Astropy},
          \texttt{emcee} \citep{emcee}, 
          \texttt{corner} \citep{corner}, 
          \texttt{vaex} \citep{vaex2018},
          \texttt{SciPy} \citep{2020SciPy-NMeth},
          \texttt{matplotlib} \citep{matplotlib},
          \texttt{NumPy} \citep{Numpy},
          \texttt{AGAMA} \citep{AGAMA}, 
          \texttt{SuperFreq} \citep{SuperFreq},
          \texttt{pyGadgetReader} \citep{PyGadgetreader} and Jupyter Notebooks \citep{JupyterNotebook}.
\end{acknowledgements}

\bibliography{bibliography}
\bibliographystyle{aa} 

\begin{appendix}

\section{Our fiducial Galactic Potential model}
\label{app:pot}

\begin{figure}[t!]
\centering
    \includegraphics[width=\hsize]{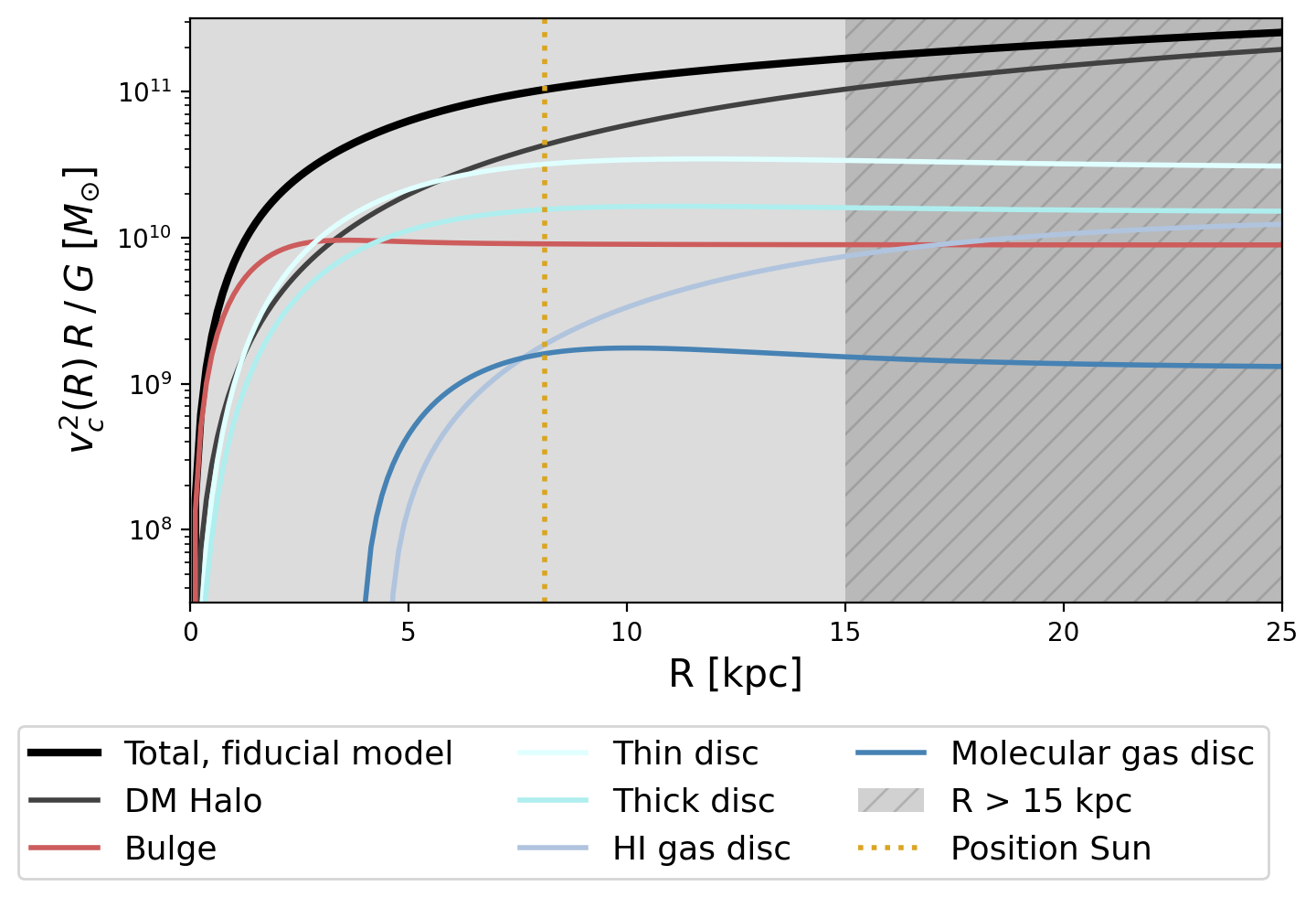}
\caption{$\dfrac{v_{\rm{c}}^2(R) \: R } {G}$ for the fiducial potential model and its different components individually. This quantity can be taken as a proxy for the mass enclosed within a certain radius $R$, and this plot thus gives a feeling for the relative importance of each component at different radii on the midplane. The orbits of the HS stars reach between 5~-~20 kpc (see Fig.~\ref{fig:orbparams}), and in this range the DM halo dominates. The grey hatched area marks the region $R > 15$~kpc and data in this distance range was not included in the MCMC fit.
\label{fig:relative_Mr}}
\end{figure}

Throughout this work, we use a modified version of the \cite{McMillan2017} potential. The \cite{McMillan2017} potential is an axisymmetric potential model consisting of a bulge, an exponential thin and thick disc, a HI gas disc, a molecular gas disc and a spherical NFW DM halo. Since recent work has shown that this model might be slightly too heavy \citep[e.g.][]{Eilers2019, Jiao2023}, we perform a simple Monte Carlo Markov Chain using \texttt{emcee} \cite{emcee} with 40 walkers and a 1000 steps to optimise the scale radius of the halo $r_{\rm{h}}$, the density of the halo $\rho_{\rm{0, h}}$ and $M_{\rm{discs}}$, the sum of the thin and thick disc mass, assuming a thick disc scale length of $R_{\rm{thick}} = 2$~kpc \citep{Bland-Hawthorn2016} and a fixed local density normalisation $f_{\rm{d}, \odot} = \rho_{\rm{thick}}(R_{\odot}, z_{\odot})/\rho_{\rm{thin}}(R_{\odot}, z_{\odot}) = 0.12$ \citep{McMillan2017, Ibata2023}. Furthermore, following \cite{Dodd2022}, we assume an axisymmetric halo with a density flattening $q = 1.2$. We use rotation curve data with associated uncertainties derived using axisymmetric Jeans equations by \cite{Eilers2019}\footnote{In the case of \cite{Eilers2019}, we take the mean of the measured uncertainties and add the systematic uncertainties (extracted from their Fig.~4), assuming a systematic uncertainty of 12\% for the five bins largest in $R$, quadratrically, such that $\sigma_{\rm{v_c, tot}}^2 = \sigma_{\rm{meas}}^2 + \sigma_{\rm{syst}}^2$. We treat the different contributions to the total systematic uncertainties as independent and add them quadratically.}, who used \textit{Gaia} DR2 data, and \cite{Zhou2023}\footnote{In the case of \cite{Zhou2023}, we similarly add the systematic uncertainties (extracted from their Fig.~12) quadratically and also treat the different contributions to the total systematic uncertainty as independent by summing them quadratically. }, who used APOGEE, LAMOST, 2MASS and \textit{Gaia} EDR3 data and employed supervised machine-learning to obtain distances\footnote{We use these two datasets as they use consistent values for $R_{\odot}$, $v_{\odot}$, and $z_{\odot}$. Alternative works on rotation curves make different assumptions on their solar motion and position. For example, \cite{Wang2023RC} assumes $R_{\odot} = 8.34$~kpc and $z_{\odot} = 0.027$~kpc, and \cite{Ou2024} assumes $R_{\odot} = 8.178$~kpc and $z_{\odot} = 0.0208$~kpc. This in turn leads to different adopted values of the solar motion.}. 
We restrict ourselves conservatively to rotation curve data with $R < 15$ kpc, because beyond this radius systematic errors associated to assumptions behind the methods used to derive the rotation curve become important \citep{KoopRC}. However, we checked that if we take data within $R <20$~kpc, our conclusions remain similar. We set a simple flat prior and allow 
\begin{equation}
P(\vec{\theta}) = \left\{
    \begin{array}{ll}
        1 & \text{if} \left\{
            \begin{array}{ll}
                10 \leq r_{\rm{h}} \leq 30 \: \text{kpc} \\
                10^{5} \leq \rho_{\rm{0, h}} \leq 10^{8} \: M_{\odot} \text{~kpc}^{-3} \\
                3 \cdot 10^{10} \leq M_{\rm{discs}} \leq 5.5 \cdot 10^{10}~M_{\odot}
            \end{array}
            \right.  \\
        0 & \text{otherwise.}
    \end{array}
\right.
\label{eq:prior}
\end{equation}
We find $v_{\rm{c}}(R_{\odot}) = 233.2 \pm 2.6 \si{\: km \: s^{-1}}$ and median and percentiles  $r_{\rm{h}} = 15.7^{+8.8}_{-4.4}$~kpc, $\rho_{\rm{h, 0}} = 10^{+8}_{-5} \cdot 10^6 \: M_{\odot} \: \text{kpc}^{-3}$ and $M_{\rm{discs}} = 4.5^{+0.6}_{-0.7} \cdot 10^{10}~M_{\odot}$, and use these median values throughout this work to set our fiducial potential. The contributions as a function of radius of all components of the potential are shown in Fig.~\ref{fig:relative_Mr}. The total mass within 20 kpc, the region probed by the Helmi Streams, of this model is $M_{\rm{tot}}(r < 20 \: \text{kpc}) = 2.2^{+0.1}_{-0.1} \cdot 10^{11} M_{\odot}$, compatible with other estimates \citep[e.g.][]{Kupper2015, Posti2019, Watkins2019, Eadie2019}.

\section{Orbital frequencies determination}
\label{app:freqsmethod}

\begin{figure*}[htb!]
\centering
    \includegraphics[width=1\textwidth]{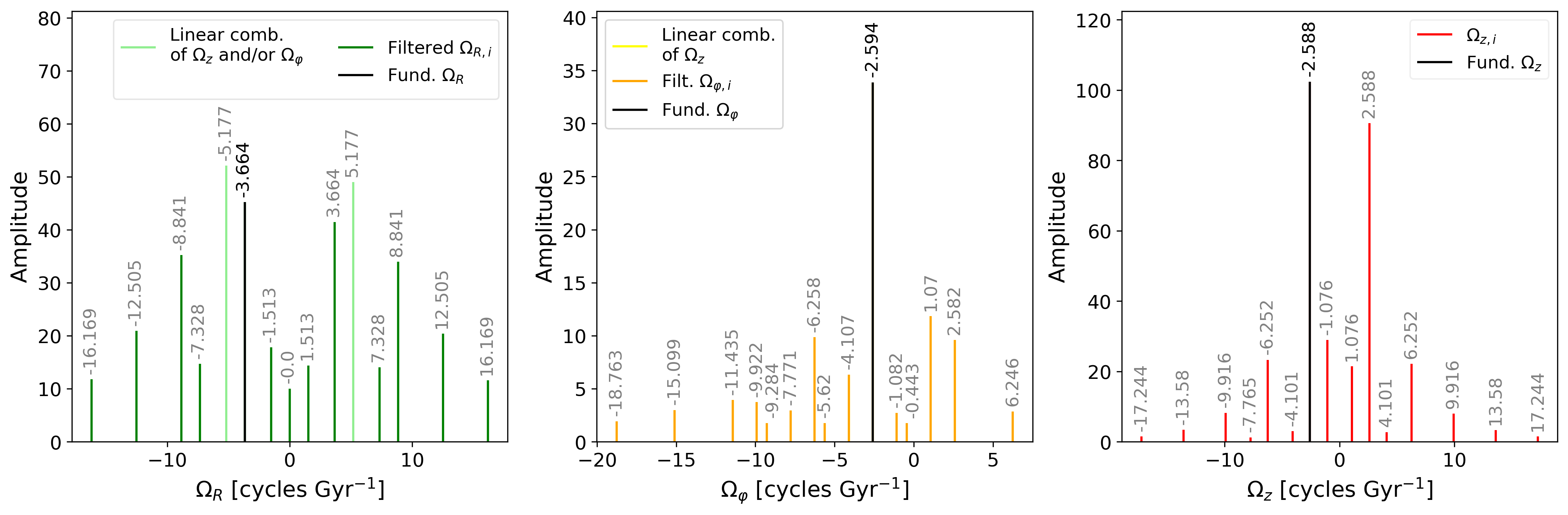}
\caption{Frequency spectra and determined fundamental (fund.) orbital frequencies of the orbit of a randomly chosen Helmi Streams' star in the fiducial potential (see Appendix~\ref{app:pot}) for an integration time of 100~Gyr, outputted each 10 Myr. The filtered (filt.) spectra for $\Omega_R$ ($\Omega_{\phi}$) correspond to spectra out of which the multiples of $\Omega_z$ and $\Omega_{\phi}$ ($\Omega_z$) are filtered. For $\Omega_R$, we see that the leading frequency of the spectrum is twice $\Omega_z$. However, the orbit's phase-space distribution does not suggest it is a resonant orbit, and this value for $\Omega_R$ does not fall within the range \texttt{abs}($\Omega_{R}: \Omega_z - 0.7$) < 0.15), and it is valid to choose the next-highest peak as fundamental $\Omega_R$.
\label{fig:freqsstar}}
\end{figure*}

To recover the fundamental orbital frequencies corresponding to a star's orbit, we use a modified version of \texttt{SuperFreq} \citep{SuperFreq}. \texttt{SuperFreq} is an implementation of the Numerical Analysis of Fundamental Frequencies (NAFF) method \citep{Laskar1993, Valluri1998}. It performs a Fast Fourier Transform of a complex time series $w(t) + i v_w(t)$, where $w$ are the coordinates, for example Poincaré symplectic coordinates \citep{Papaphilippou1996} or Cartesian coordinates. If a 3D time series is inputted, \texttt{SuperFreq} stores the three frequency spectra (frequency and amplitude) of each of the time series and sorts them by amplitude. Next, the frequency with the highest amplitude is chosen as the first leading frequency. It then determines the second leading frequency by picking the frequency with the second highest amplitude, requiring it to be of a different coordinate and requiring a minimum difference in frequency between the first two leading frequencies (this difference is by default set to $10^{-6}$ but can be changed). In a similar way, the third leading frequency is found. This set of three leading frequencies typically corresponds to the set of three fundamental frequencies, but this does not necessarily need to be the case, something that was noted by \cite{Dodd2022} and \cite{Silva2023} as well.

To make the fundamental frequency determination more robust, we proceed as follows. We integrate orbits for 100~Gyr, as the longer an orbit is integrated, the more reliable the frequency determination becomes \citep{Valluri2010, Wang2016}. We output each 10~Myr and construct a set of three time series in Poincaré symplectic polar coordinates for each orbit\footnote{Poincaré's symplectic polar coordinates, a variation on cylindrical polar coordinates, are used as input for \texttt{SuperFreq} throughout this work, and thus also in the case of mildly triaxial potentials. This choice of coordinates ensures relatively good results and gives more information about the orbital properties than for example the orbital frequencies in Cartesian coordinates do.}. This is used as input for \texttt{SuperFreq}, which determines the frequency spectra with settings \texttt{Nintvec} $= 15$ and \texttt{break condition = None}. \footnote{To a certain extent, the frequencies recovered by the algorithm are dependent on the settings in the code, as also shown by \cite{Wang2016}.}. We start with the $z$ coordinate, and from its frequency spectrum define the frequency with the highest amplitude to be $\Omega_z$. Next, we filter the frequency spectra of $R$ and $\phi$ such that frequencies that are a multiple of $\Omega_z$ are removed. This is done by requiring a minimum difference of $10^{-4}$ between $\Omega_{\phi}$ and $\Omega_z$ and $\Omega_R$. Of course, this presents issues if the orbit in question is a resonant orbit. Therefore, we step wise make the required minimum difference smaller ($10^{-5}$, $10^{-6}$, $10^{-7}$, $10^{-8}$) if the peak that is picked results in an unexpected frequency ratio. For the Helmi Streams, this translates to the requirement \texttt{abs}($\Omega_{\phi}: \Omega_z - 1$) < 0.15, which effectively means that we assume that all Helmi Stream stars have orbital frequencies that are drawn from a continuous distribution, i.e. they have values that are within a certain range. Next, the frequency with the highest amplitude in the $\phi$ spectrum is chosen as $\Omega_{\phi}$. Then, the remaining frequency spectrum of $R$ is filtered such that the $R$ frequencies that are a multiples of $\Omega_{\phi}$ are removed. Finally, the frequency with highest amplitude that is independent is chosen as $\Omega_R$. However, as linear integer combinations of this peak with $\Omega_z$ or $\Omega_{\phi}$ may appear, one needs to be careful to choose the peak that corresponds to the independent $\Omega_R$. For the Helmi Streams, this is done by requiring that \texttt{abs}($\Omega_{R}: \Omega_z - 0.7$) < 0.15). The fact that multiples of the peaks in $\Omega_z$, $\Omega_{\phi}$ and $\Omega_R$ appear in all three spectra is indicative of the fact that the motion in $R$, $\phi$ and $z$ is coupled. The procedure to determine the orbital frequencies is illustrated in Fig.~\ref{fig:freqsstar} and solves the problem of the ficticious $\Omega_z : \Omega_R = 1:2$ branch reported by \cite{Dodd2022}.

\section{Orbit maps of a Helmi Streams'  subclump star}
\label{appendix:maps}

\begin{figure*}[htb!]
\centering
    \includegraphics[width=\hsize]{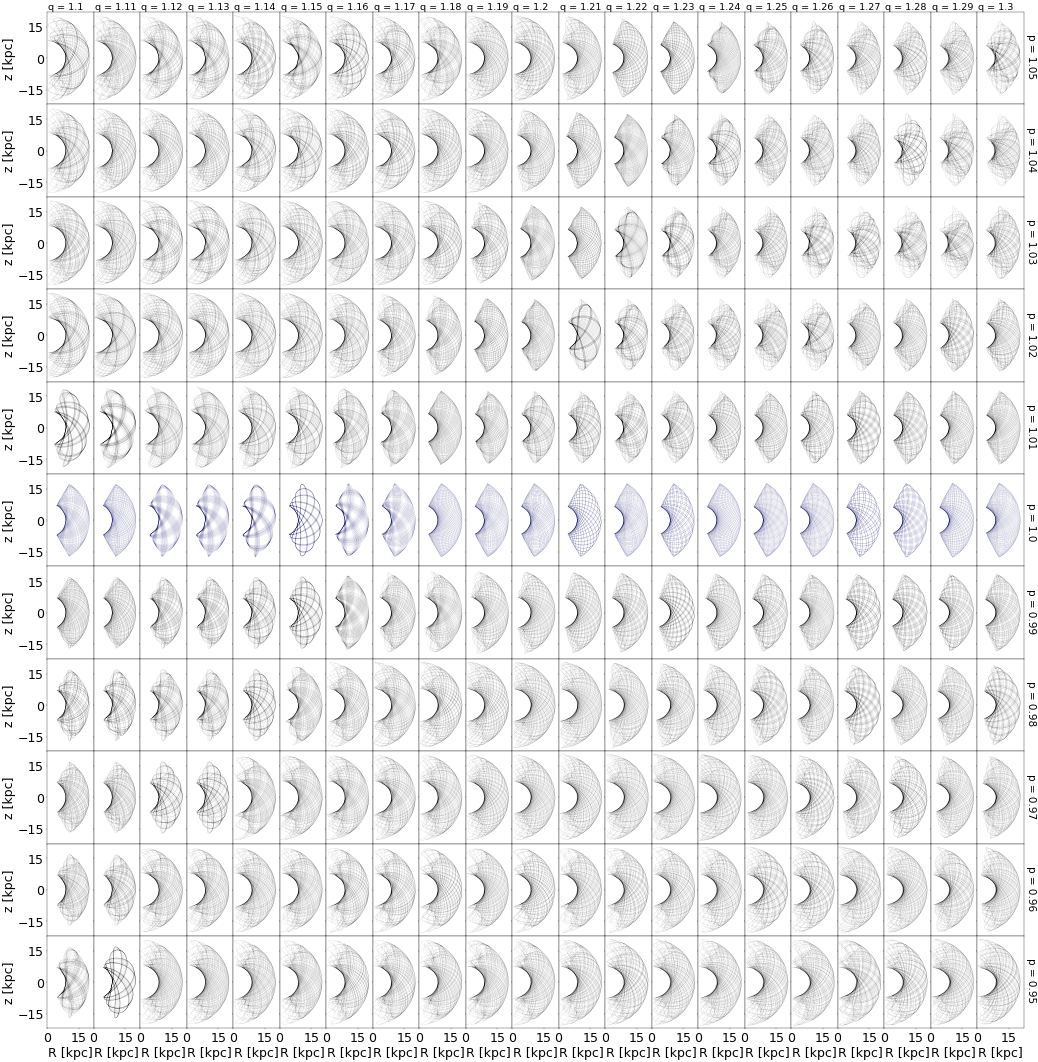}
\caption{Orbit in Galactocentric cylindrical ($R, z$) of a Helmi Streams' subclump star in the fiducial potential  (see Appendix~\ref{app:pot}) with a triaxial NFW halo for a range of $p$ and $q$. Orbits in axisymmetric potentials are indicated in dark blue. 
\label{fig:orbitmapRz}}
\end{figure*}

\begin{landscape}
\begin{figure}[h!]
\centering
    \includegraphics [width=\linewidth,keepaspectratio]{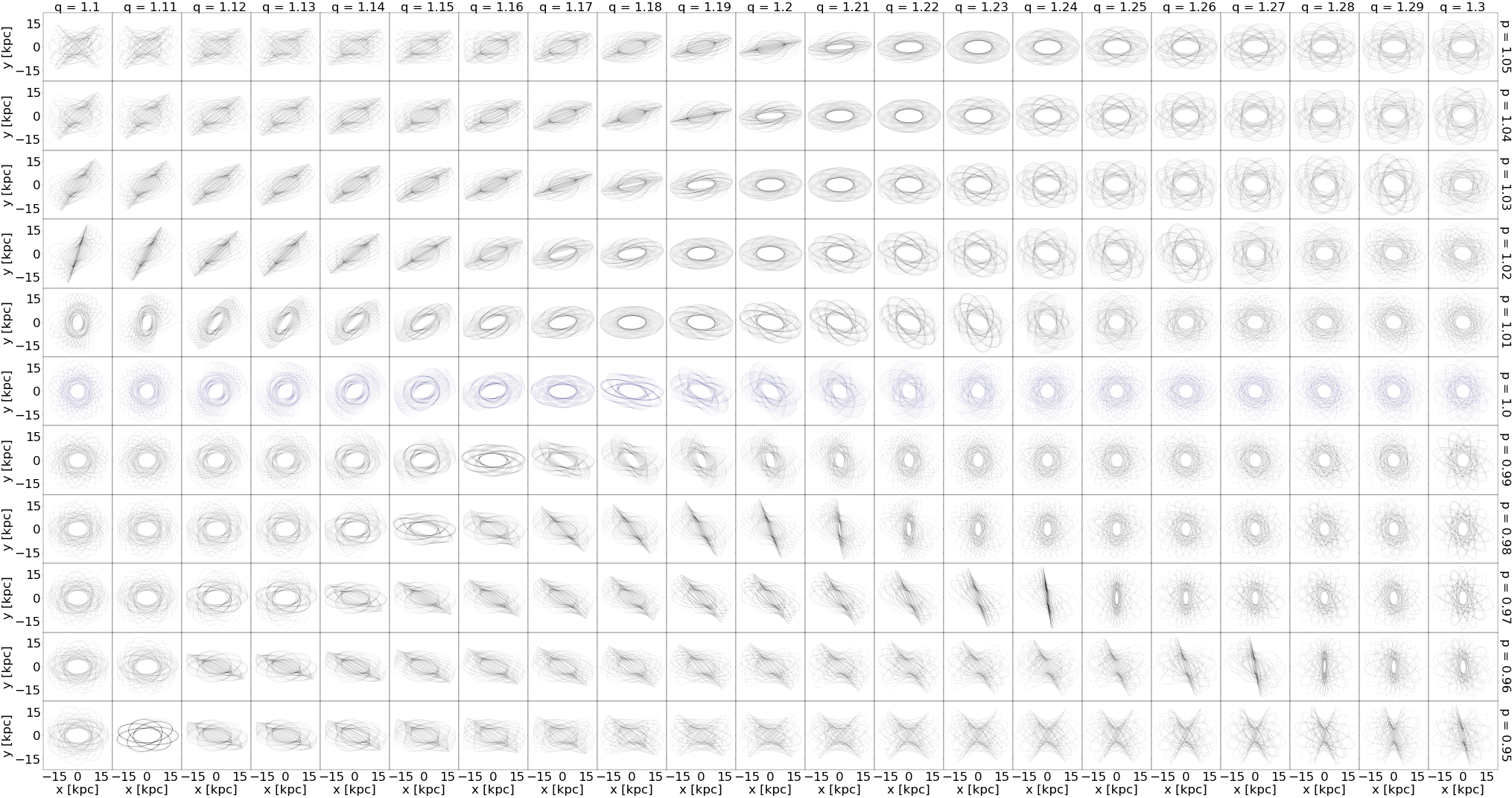}
\caption{Orbit in Galactocentric Cartesian ($x, y$)  of a Helmi Streams' subclump star in Galactic potentials with a triaxial NFW halo for a range of $p$ and $q$. Orbits in axisymmetric potentials are indicated in dark blue. 
\label{fig:orbitmapxy}}
\end{figure}
\end{landscape}

\begin{landscape}
\begin{figure}[h!]
\centering
    \includegraphics[width=\linewidth,keepaspectratio]{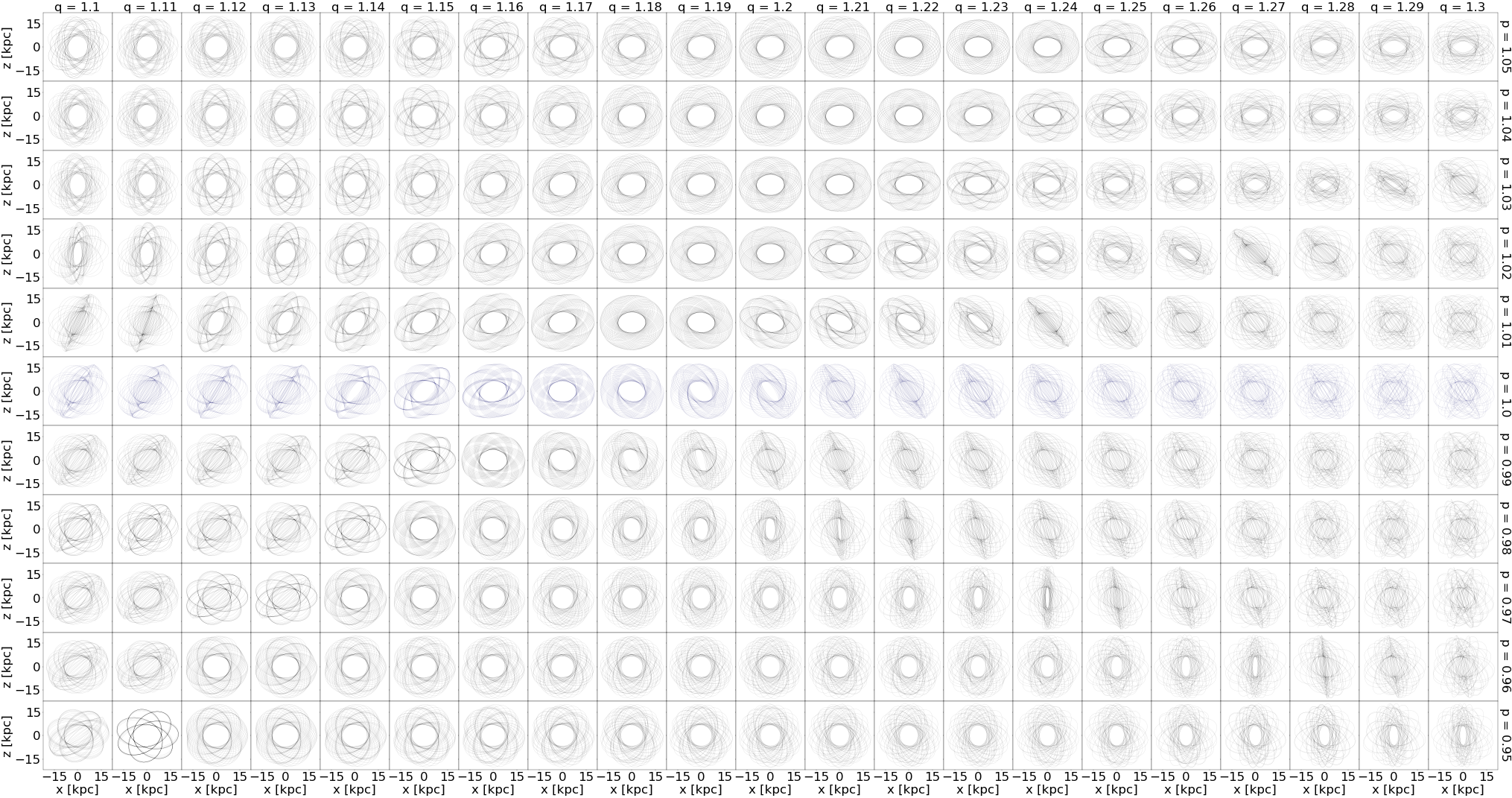}
\caption{ Orbit in Galactocentric Cartesian ($x, z$)  of a Helmi Streams' subclump star in Galactic potentials with a triaxial NFW halo for a range of $p$ and $q$. Orbits in axisymmetric potentials are indicated in dark blue. 
\label{fig:orbitmapxz}}
\end{figure}
\end{landscape}

\begin{landscape}
\begin{figure}[h!]
\centering
    \includegraphics[width=\linewidth,keepaspectratio]{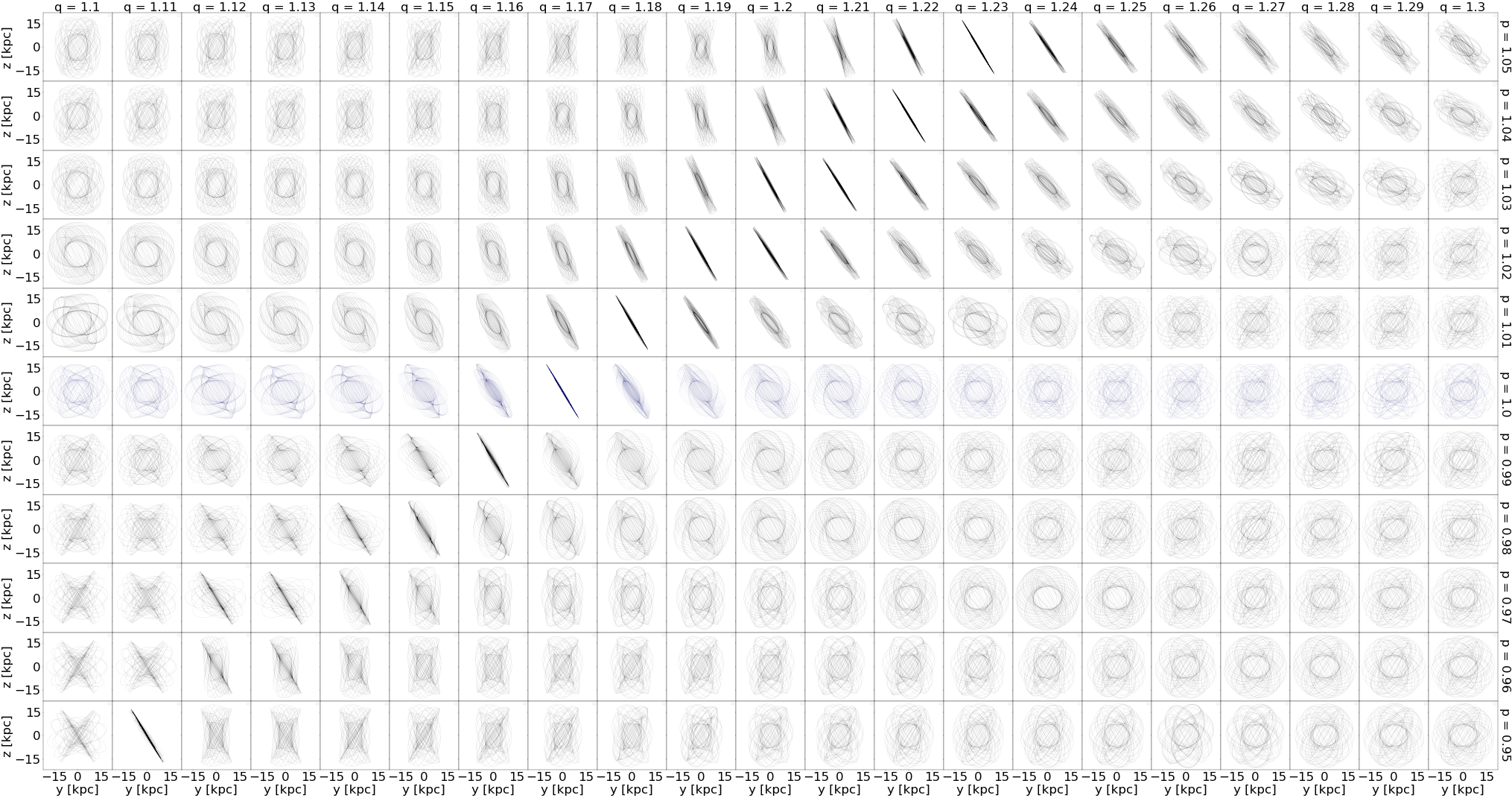}
\caption{Orbit in Galactocentric Cartesian ($y, z$) of a Helmi Streams' subclump star in Galactic potentials with a triaxial NFW halo for a range of $p$ and $q$. Orbits in axisymmetric potentials are indicated in dark blue. The $\Omega_{\phi}: \Omega_z$ is 1:1 resonance stands out clearly, as the orbit becomes planar. 
\label{fig:orbitmapyz}}
\end{figure}
\end{landscape}

We randomly picked a subclump star and integrated its orbit for 20~Gyr in the fiducial potential with a triaxial NFW halo for a range of $0.95 < p < 1.05$ and $1.10 < q < 1.30$ to create orbit maps. Figure~\ref{fig:orbitmapRz} shows the map in $(R, z)$, Fig.~\ref{fig:orbitmapxy} shows the map in $(x, y)$, Fig.~\ref{fig:orbitmapxz} shows the map in $(x,z)$ and Fig.~\ref{fig:orbitmapyz} shows the map in $(y,z)$. By inspecting these maps, different orbit families can be recognised. The resonantly trapping $\Omega_{\phi}: \Omega_z$ is 1:1 resonance clearly stands out for $p>1$ in Fig.~\ref{fig:orbitmapyz}, while such a region is not apparent for $p <1$. The $y$-tubes and $z$-tubes can be recognised easily from  Fig.~\ref{fig:orbitmapxz} and Fig.~\ref{fig:orbitmapyz}, as on these families the star orbits rotate around the $y$ and $z$ axis, respectively.

\section{Orbital frequencies of the Helmi Streams}
\label{app:freqs}

\begin{figure*}[htb!]
\centering
    \includegraphics[width=\hsize]{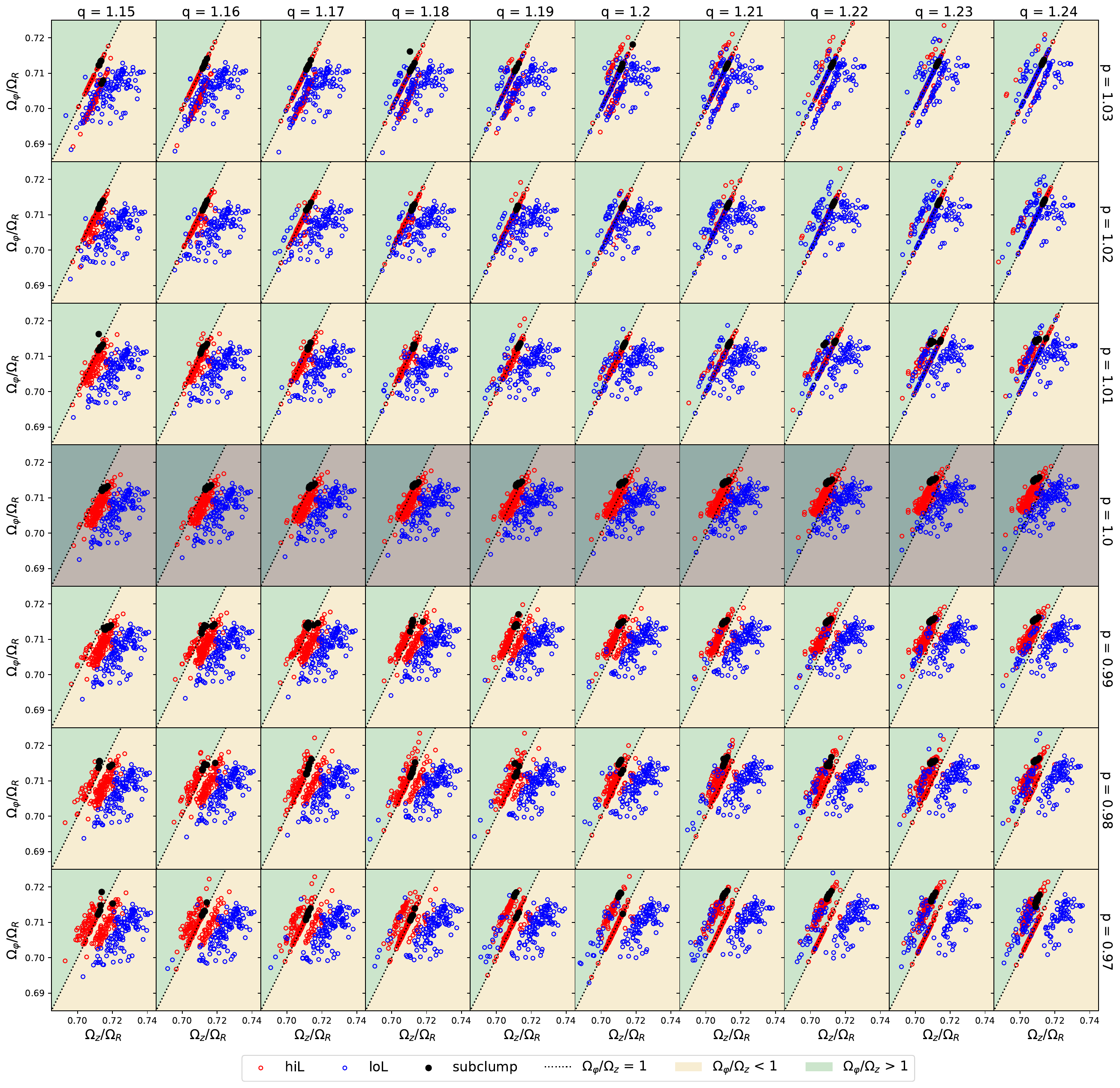} 
\caption{Orbital frequency distribution  of the Helmi Streams' stars in Galactic potentials with a triaxial NFW halo for a range of $p$ and $q$. Axisymmetric potentials ($p =1$) are indicated with a blue-grey facecolor. The hiL stars are encircled in red, the loL stars in blue and the subclump stars in black. The dotted line indicates the $\Omega_{\phi} : \Omega_z$ 1:1 resonance, which acts as a separatrix in potentials with $p <1$ and resonantly traps orbits for $p >1$.
\label{fig:HS_freqspace}}
\end{figure*}

Figure~\ref{fig:HS_freqspace} shows the distribution of orbital frequencies of the Helmi Streams stars in the potential where we have replaced the DM halo with a triaxial halo for a range of $p$ and $q$. If stars are on the $\Omega_{\phi} : \Omega_z$ resonance, they fall on the dotted line in the figure. Using $\Delta L_y$ and $\Delta L_z$ to classify the orbits, we checked and found that the orbital frequencies of $z$ tube orbits and truly resonant orbit are determined reliably. However, the algorithm sometimes has some problems to  determine the orbital frequencies of $y$ tube orbits, which are either classified as resonant orbits or placed on a wrong branch, which both stand out as line features in Fig.~\ref{fig:HS_freqspace}. This can be attributed to the fact that the cylindrical coordinates $(R, \phi, z)$ are not the most suitable coordinates to describe the motion of $y$ tube orbits. However, the frequency determination is good enough for our purposes, which is to use the median of the orbital frequencies of the hiL, loL and subclump stars to map out where orbital resonances reside (see Fig.~\ref{fig:mapstogether} of the main text). It is also interesting to see the effect of the $\Omega_{\phi} : \Omega_z$ 1:1 resonance for $p >1$, clearly trapping orbits, and $p < 1$, resulting in a chaotic and thus underpopulated region in frequency space.

\section{\texttt{Progenitor 4} in potentials with different ($p, q$)}
\label{appendix:sims_pq}

\begin{landscape}
\begin{figure}[t!]
\centering
    \includegraphics[width=\hsize]{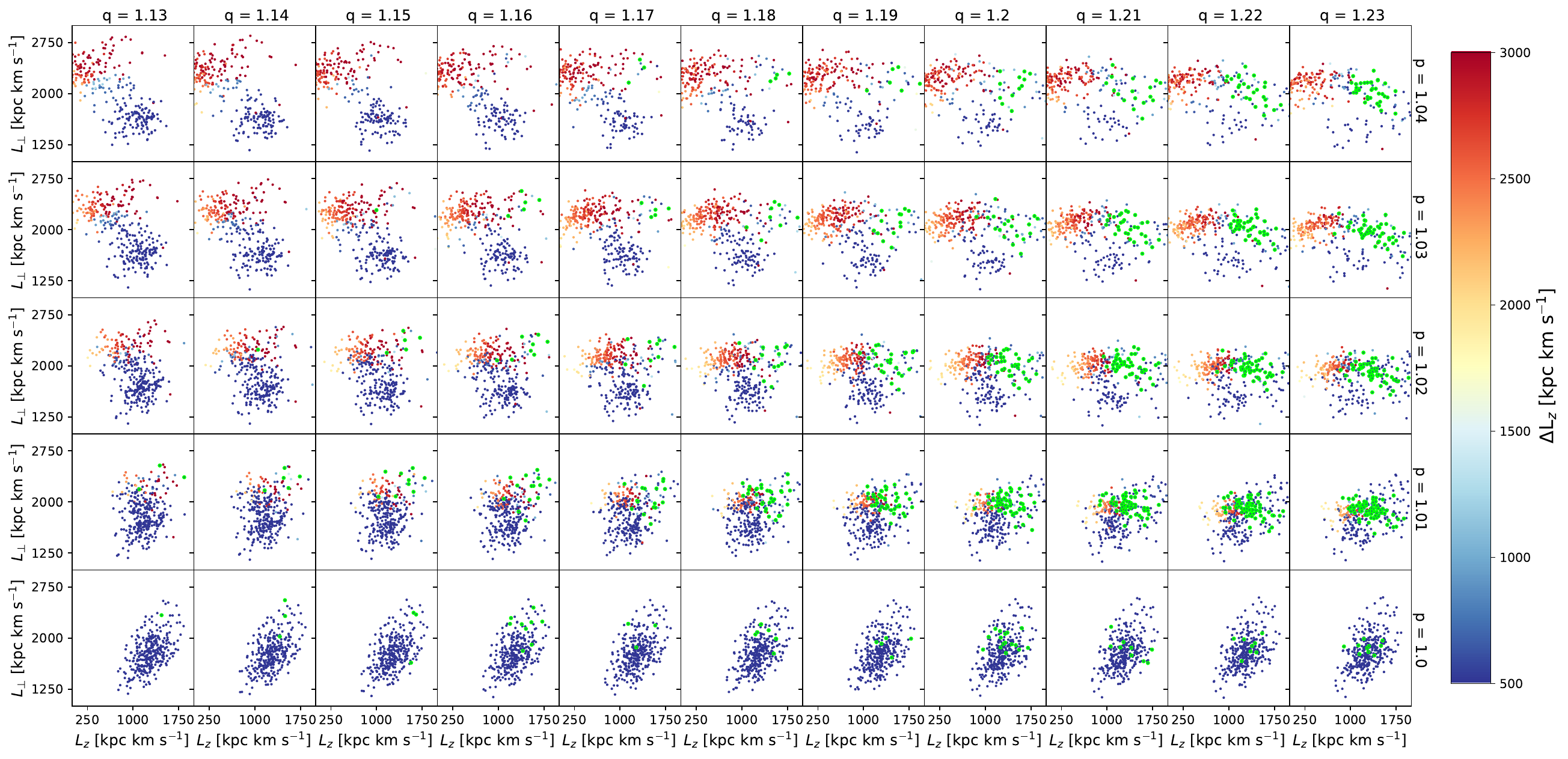}
\caption{Snapshot of the ($L_z$, $L_{\bot}$) distribution after  $8$~Gyr of integration time of 319 particles with Helmi-Stream-like phase-space positions, selected from the re-centred \texttt{Progenitor 4} by \cite{Koppelman2019Helmi}, in Galactic potentials with a triaxial NFW halo for a range of $p$ and $q$. The colourbar indicates the variation in $L_z$ over an integration time of 100~Gyr. In the axisymmetric potential, $p =1.00$, $L_z$ is an IoM and all particles are on $z$-axis tube orbits. For $p > 1$, the particles with larger $L_{\bot}$ are on $y$-axis tube orbits, meaning their $L_z$ is no longer conserved, making them move towards lower $L_z$ in time with respect to the particles on $z$-axis tube orbits. The timescale over which this happens depends on the degree of triaxiality of the potential, and in particular the value of $p$. A number of particles are strongly trapped by the $\Omega_{\phi}:\Omega_z$ 1:1 resonance (selected as $\Delta L_{y,z} \lesssim 750$), and are encircled in green. A video, showing the behaviour of the particles in $(L_z, L_{\bot})$ space over time, can be found \href{https://hannekewoudenberg.github.io/helmi_streams/}{here}.
\label{fig:ICHS_all}}
\end{figure}
\end{landscape}

In Sect.~\ref{sec:analysis:sims} we explore the ($L_z$,~$L_{\bot}$) distributions of \texttt{Progenitor 4} particles after they have have been re-centred to the 6D phase-space coordinates of a central HS star 8~Gyr ago and integrated forward to the present day. Figure~\ref{fig:ICHS_all} shows  the  ($L_z$,~$L_{\bot}$) distribution after 8~Gyr of integration time for the range $1.13 < q < 1.23$ and $1.00 < p < 1.04$. For the range $1.18 < q < 1.21 $ and $1.02 < p < 1.03$ we find distributions that match the HS' observations most convincingly, showing two separated clumps of stars, driven by the fact that the stars in the two clumps are on different orbital families.

\end{appendix}

\end{document}